\documentclass[a4paper,fleqn]{cas-sc}

\usepackage[numbers]{natbib}
\usepackage{xcolor}
\usepackage{graphicx}
\usepackage{dcolumn}
\usepackage{bm}
\usepackage{subcaption}
\usepackage{siunitx}
\usepackage{multirow}
\usepackage{amssymb}
\usepackage{hyperref}
\usepackage{booktabs}
\usepackage{tabularx}
\usepackage{array}
\usepackage{makecell}
\usepackage{xurl}
\usepackage{lineno}
\usepackage[normalem]{ulem}

\def\tsc#1{\csdef{#1}{\textsc{\lowercase{#1}}\xspace}}
\tsc{WGM}
\tsc{QE}

\begin{document}
\let\WriteBookmarks\relax
\def\floatpagepagefraction{1}
\def\textpagefraction{.001}

\shorttitle{}    

\shortauthors{}  


\title[mode=title]{Optical effects in Gaseous Electron Multipliers (GEMs)}{}  



%

\author[inst1]{D. Edgeman}[orcid=0009-0002-8836-3801]
\cormark[1]
\ead{dedgeman2023@unm.edu}

\author[inst2]{F. M. Brunbauer}
\author[inst1]{M. Gardner}
\author[inst1]{D. Loomba}
\author[inst3,inst5]{P. A. Majewski}
\author[inst3]{T. Marley}
\author[inst4,inst5]{L. Millins}
\author[inst4]{T. Neep}
\author[inst4,inst6]{K. Nikolopoulos}
\author[inst1]{J. Schueler}
\author[inst1]{E. Tilly}
\author[inst1]{W. Thompson}

\address[inst1]{Department of Physics and Astronomy, University of New Mexico, Albuquerque, NM 87131, USA}
\address[inst2]{CERN, 1211 Geneva 23, Switzerland}
\address[inst3]{Department of Physics, Blackett Laboratory, Imperial College London, London, SW7 2AZ, UK}
\address[inst4]{School of Physics and Astronomy, University of Birmingham, Birmingham, B15 2TT, UK}
\address[inst5]{Particle Physics Department, STFC Rutherford Appleton Laboratory, Didcot, OX11 0QX, UK}
\address[inst6]{University of Hamburg, 22767 Hamburg, Germany}


\cortext[1]{Corresponding author}



\begin{abstract}
Optical time projection chambers (OTPCs) are well suited for applications that require the highest spatial resolution for particle track reconstruction. The MIGDAL experiment uses a glass GEM-based OTPC and observes a systematic excess in both the intensity and width of particle tracks in its optical readout, when compared with charge readout simulations. One hypothesis is that scintillation light produced inside a GEM hole during the avalanche propagates through the GEM substrate and exits neighboring holes. We present lab measurements testing this hypothesized optical broadening effect in three types of GEM substrates: glass, ceramic, and FR4. Our observations quantify this optical broadening and demonstrate it to be strongest in glass GEMs. Additionally, we use Geant4 simulations to both reproduce our observations and quantify optical broadening effects in realistic charge avalanches. Applying our glass GEM effects to simulated particle tracks yields increases of track intensity and widths by up to approximately $26$\% and $31$\%, respectively. This may explain the larger-than-expected track intensities and widths observed in the MIGDAL OTPC. This effect may be relevant to GEM-based OTPCs, with its magnitude depending on the GEM structure and detector-specific optical configuration.
\end{abstract}





\maketitle

\section{\label{sec:Intro}Introduction}
Since its invention 50 years ago \cite{Nygren1978}, the time projection chamber (TPC) has become a workhorse in experiments used in particle and nuclear physics. Its ability to measure low energy particle tracks with high resolution in three dimensions makes it an ideal detector for rare event searches, which require the use of directionality to distinguish signals from background \cite{Vahsen2021}. These include searches for dark matter \cite{Morgan2003, Battat2009, Shimada2023, Santos_2011, Amaro_2023}, rare nuclear decays \cite{Wang2020, Sokolowsak2024}, X-ray polarimetry \cite{Black2007, Fiorina2024}, and the Migdal effect \cite{MIGDAL2023, Xu2024}. 

TPCs span a large range of sizes, from tabletop to $\mathcal{O}(100)$m$^3$ \cite{Alme2010}, and have great versatility in the choice of medium -- whether it is liquid or gas -- and on technologies used to amplify and read out their signals. The latter can be divided into two broad classes: charge and optical readouts. These measure ionization and scintillation, respectively, produced during amplification of the primary signal. Due to rapid advances in technology, the use of optical readout TPCs (OTPCs) has seen a rapid growth over the past decade \cite{FRAGA2001125,PINCI2019453,Brunbauer2018,Brunbauer2018-2,Phan2020,Roberts-2019,Lowe-2020,Antonietti-2025}. These advances are due to the development of micro-pattern gas detectors (MPGDs) used to amplify the charge and light signals, and scientific grade camera technologies used to detect the latter. We describe each below, emphasizing those features that provide unique advantages of an OTPC.

The advances in scientific grade charge-coupled device (CCD) and complementary metal-oxide-semiconductor (CMOS) camera technologies have largely been driven by industry. The current state-of-the-art cameras boast sufficiently high quantum efficiencies\footnote{Up to $90\%$ in relevant wavelengths spanning near ultraviolet (UV), optical and near
IR.} and low readout and dark current noise necessary to detect and resolve single photons. High granularity sensors with a large number of small pixels ($\gtrsim64$ Mpix\footnote{https://specinstcameras.com/vision-64/}) together with tunable optical coupling with lenses and mirrors provide the flexibility to image larger areas without sacrificing spatial resolution. 
Having an off-the-shelf camera-lens readout that is decoupled from the TPC is a cost-effective solution leading to much shorter development times.

Decoupling the camera-lens readout system from the TPC also makes it insensitive to electronic noise from the detector, reducing labor and development times. Together with the low cost per ``channel", i.e. pixel, of the camera, all these features make the OTPC well suited for detector R\&D and for the rapid deployment of cost-effective experiments. One drawback is that full 3D tracking typically requires a charge readout or auxiliary fast photon detectors to reconstruct the track component along the drift axis due to the limited frame rate of imaging sensors \cite{MIGDAL2023,Brunbauer2018-2}. This will soon be obviated as camera readout speeds are approaching and may eventually exceed those required to do this optically \cite{BrunbauerRD51_2018}. 

Advances in MPGDs have kept apace and complemented those in camera technologies. In OTPCs, one of the roles of these devices is to provide gas amplification that results in the scintillation signal read by the camera. Although both gas electron multipliers (GEMs) and Micromegas have been used for this purpose in OTPCs \cite{Abe2011}, we focus on the former in this paper. Two main features of GEMs are fine pitch and high gas gains across a wide range of pressures. Fine pitch, which couples well to the small pixels of CCD and CMOS cameras, enables high-definition imaging of short, low-energy tracks. This, in turn, results in many independent samples of the track for use in precise track reconstruction. High gain is critical to achieving the light yields needed for the high signal-to-noise (S/N) detection of faint, low ionization-loss (d$E$/d$x$) particle tracks, as required by many applications. The ability to stack GEMs to achieve the necessary gain is another advantage of MPGDs over multi-wire proportional chambers (MWPCs) and other similar technologies.

These advantages make GEM-based OTPCs especially well suited for applications that require high spatial resolution for particle track reconstruction. Since OTPCs measure the scintillation light produced in the secondary electron avalanche, an assumption is implicitly made that the 
information derived from the track using light is an accurate proxy for that using charge. That is, the reconstruction of the track properties (energy, d$E$/d$x$, topology, etc) derived using light faithfully reproduces those reconstructed using charge. In this paper, we experimentally test this assumption to provide a better understanding of any underlying effects involved in the optical readout approach.

The validity of the assumption that light traces charge came into question in the MIGDAL experiment \cite{MIGDAL2023}, which employs an OTPC with a Hamamatsu ORCA-Quest CMOS camera imaging the bottom of a double-GEM device made from PEG3 glass \cite{TAKAHASHI2013} through an EHD-25085-C F0.85 lens \cite{ehdlens}. It was noted that the optically imaged tracks of high d$E$/d$x$ nuclear recoils exhibited larger apparent areas than those simulated from the charge avalanche \cite{Schueler2025}. One hypothesis to explain this effect is that some of the secondary scintillation produced in a given GEM hole passes into the glass substrate and out from neighboring holes. The net effect is i) an optical ``halo" extending beyond the track defined by the charge distribution and, ii) a higher intensity of the overall track. Higher intensities are observed because a larger fraction of the isotropic light produced in a single GEM hole arrives at the camera than the fraction of the full solid angle, 4$\pi$ subtended by the hole.

Previous optical-readout studies have shown that recorded images can also be affected by detector and imaging-system response effects, including finite optical point-spread functions, lens aperture, optical aberrations, optical coupling, and reflected-light contributions \cite{Cools_2024,Amaro_2026}. In particular, Cools et al. observed reflection patterns associated with optically read out amplification structures and showed that these patterns can depend on the geometry of the camera-facing electrodes \cite{Cools_2024}. CYGNO light-response modeling connects avalanche-produced light in GEM multiplication channels to the camera signal through geometrical optical acceptance, lens transmission, and camera quantum efficiency \cite{Amaro_2026}. These studies establish that optical-response effects are important in OTPC imaging, but they do not consider 
light transport in the GEM-substrate and its effect on the track properties described above. The present work addresses this gap by studying GEM-induced optical effects in which light originating in, or reflected back toward, the GEM structure can propagate through the substrate, exit neighboring holes, and alter reconstructed track observables.

The goal of this work is to quantify the effect described above and how it manifests as differences in the observed properties of particle tracks, as seen in light versus charge. This is accomplished by studying the multi-hole point spread function (MH-PSF) resulting from light produced in a single GEM hole. Section \ref{sec:Methods} describes the experimental methods and measurements performed. Section \ref{sec:Results} details the results. Section \ref{sec:Simulations} provides a more accurate method for simulating MH-PSFs using Geant4$\,$\cite{GEANT4:2002zbu}. Section \ref{sec:Discussion} discusses how these results will impact optical GEM readouts for experiments such as MIGDAL. Discussion and summary of this work are in Section \ref{sec:Summary}.

\section{\label{sec:Methods}Measuring the MH-PSF}

In this section, we detail the equipment, experimental setup and procedure used for the study. Three different GEMs are compared: a glass GEM (G-GEM), thick GEM (THGEM), and a multi-thick GEM (M-THGEM). Standard thin GEMs were attempted but unsuccessful as described below. More details regarding the materials, dimensions, and hole geometry for each successfully tested GEM can be found in Table~\ref{tab:Results}.

\begin{table}[pos=htbp]
\caption{\label{tab:Results}Specifications and MH-PSF results for G-GEM, THGEM, and M-THGEM.}
\centering
\small
\renewcommand{\arraystretch}{1.15}
\setlength{\tabcolsep}{4pt}

\begin{tabular*}{\textwidth}{@{\extracolsep{\fill}}%
>{\raggedright\arraybackslash}p{0.26\textwidth}
>{\raggedright\arraybackslash}p{0.38\textwidth}
>{\raggedright\arraybackslash}p{0.26\textwidth}
@{}}
\toprule
\textbf{GEM materials} &
\textbf{Dimensions} &
\textbf{MH-PSF intensity at distance$^{*}$} \\
\midrule

\makecell[l]{1) Glass GEM (G-GEM)\\Cu, PEG3 \cite{TAKAHASHI2013} glass}
&
\makecell[l]{Hole diameter: \SI{170}{\micro\meter}\\
Pitch: \SI{280}{\micro\meter}\\
Thickness: \SI{570}{\micro\meter} glass, \SI{2}{\micro\meter} Cu per side}
&
\makecell[l]{1$\times$Pitch: $4 \pm 1\%$\\
2$\times$Pitch: $0.52 \pm 0.07\%$\\
3$\times$Pitch: $0.29\pm 0.03\%$}
\\
\midrule

\makecell[l]{2) Thick GEM (THGEM)\\Cu, FR4 (fiberglass)}
&
\makecell[l]{Hole diameter: \SI{300}{\micro\meter} + \SI{50}{\micro\meter} rim\\
Pitch: \SI{450}{\micro\meter}\\
Thickness: \SI{400}{\micro\meter} FR4, \SI{5}{\micro\meter} Cu per side}
&
\makecell[l]{1$\times$Pitch: $2.5 \pm 0.7\%$\\
2$\times$Pitch: $0.17 \pm 0.04\%$\\
3$\times$Pitch: $0.029 \pm 0.006\%$}
\\
\midrule

\makecell[l]{3) Multi-Thick GEM (M-THGEM)\\Cu, ceramic}
&
\makecell[l]{Hole diameter: \SI{200}{\micro\meter} + \SI{75}{\micro\meter} rim\\
Pitch: \SI{400}{\micro\meter}\\
Thickness: \SI{1600}{\micro\meter} ceramic, unknown Cu\\
3$\times$ ceramic layers, 4$\times$ electrodes (2$\times$ internal)}
&
\makecell[l]{1$\times$Pitch: $1.8 \pm 0.5\%$\\
2$\times$Pitch: $0.13 \pm 0.02\%$\\
3$\times$Pitch: $0.02 \pm 0.03^{**}\%$}
\\

\bottomrule
\end{tabular*}

\vspace{3pt}
\parbox{\textwidth}{\footnotesize
$^{*}$The MH-PSF intensity at distance lists the average intensity relative to the central peak intensity as measured by slices through GEM holes (Fig.~\ref{fig:ResultsPhase1}, right). The full 2D MH-PSF is shown in Fig.~\ref{fig:ResultsPhase1}, left and center.\\[3pt]
$^{**}$Although the relative intensity is nonzero, the uncertainty is larger than the measured value; this result is statistically consistent with zero.
}
\end{table}

To measure the MH-PSF we developed a method to produce light from inside a single isolated GEM hole. The method involves filling a single hole, as uniformly as possible, with phosphorescent paint.
This method should be interpreted as a controlled optical probe used to experimentally test our hypothesis that
scintillation in a single GEM hole passes into the glass substrate and out through neighboring holes. It is not meant to model the result from real avalanche scintillation in GEMs, whose exact spatial and wavelength distributions of light are likely to differ from that produced by the paint.
We return to the question of the detailed  profile of scintillation light produced in an avalanche in Section \ref{sec:Simulations}.

\begin{figure}[pos=htbp]
    \centering
    \begin{subfigure}[t]{0.40\textwidth}
        \centering
        \includegraphics[width=1\textwidth]{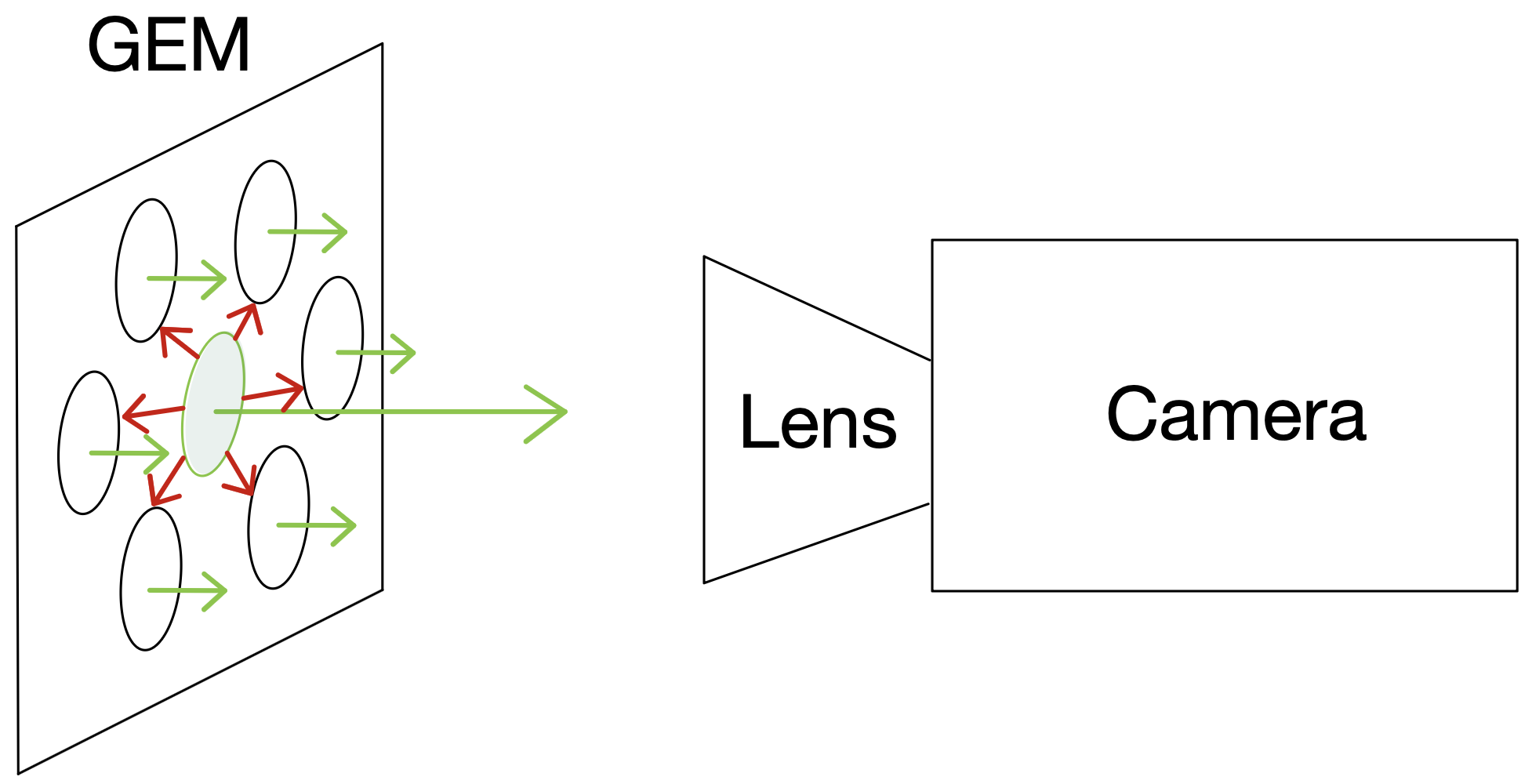}
        \caption{}
    \end{subfigure}
    \begin{subfigure}[t]{0.28\textwidth}
        \centering
        \includegraphics[width=1\textwidth]{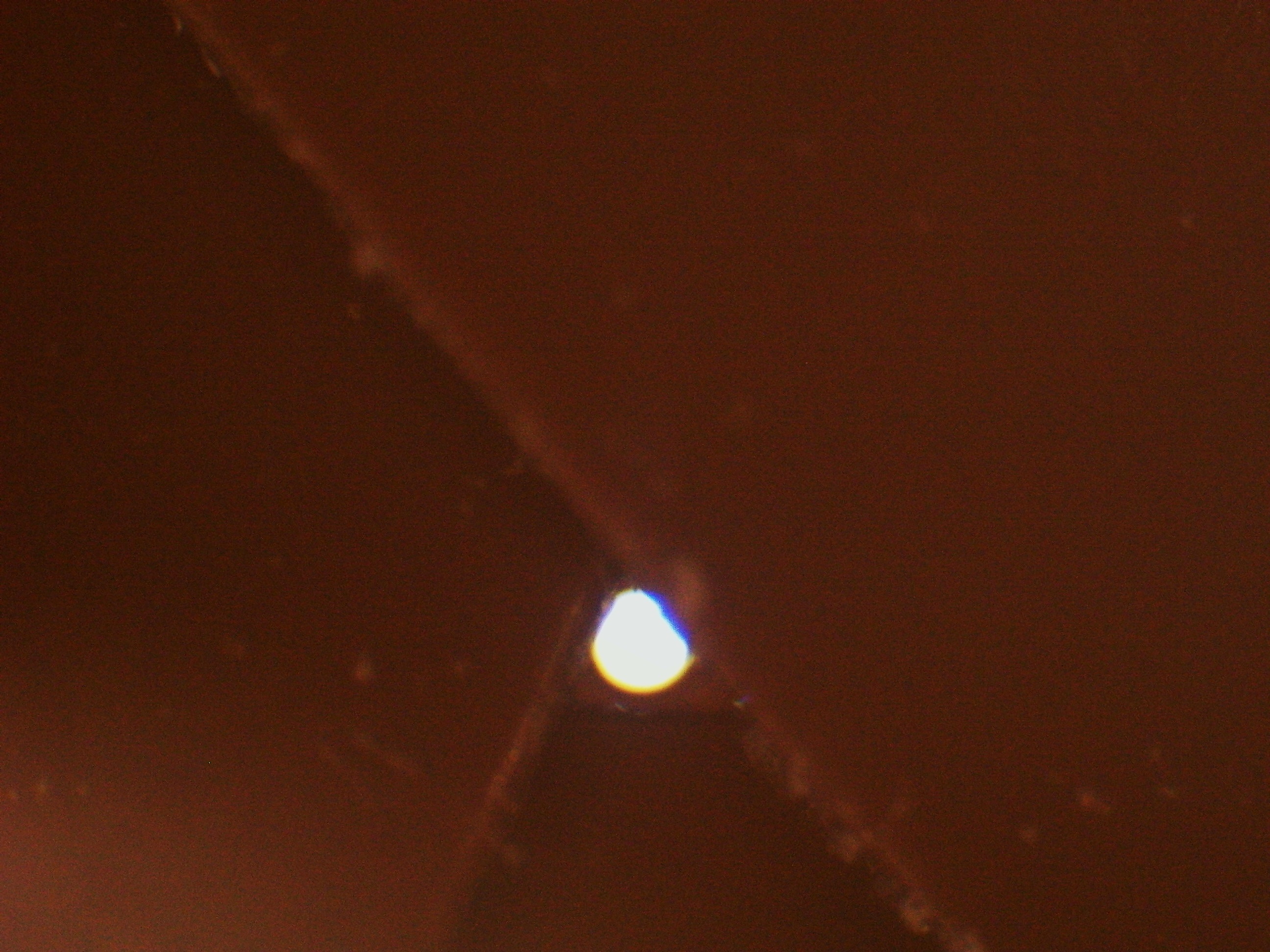}
        \caption{}
    \end{subfigure}
    \begin{subfigure}[t]{0.28\textwidth}
        \centering
        \includegraphics[width=1\textwidth]{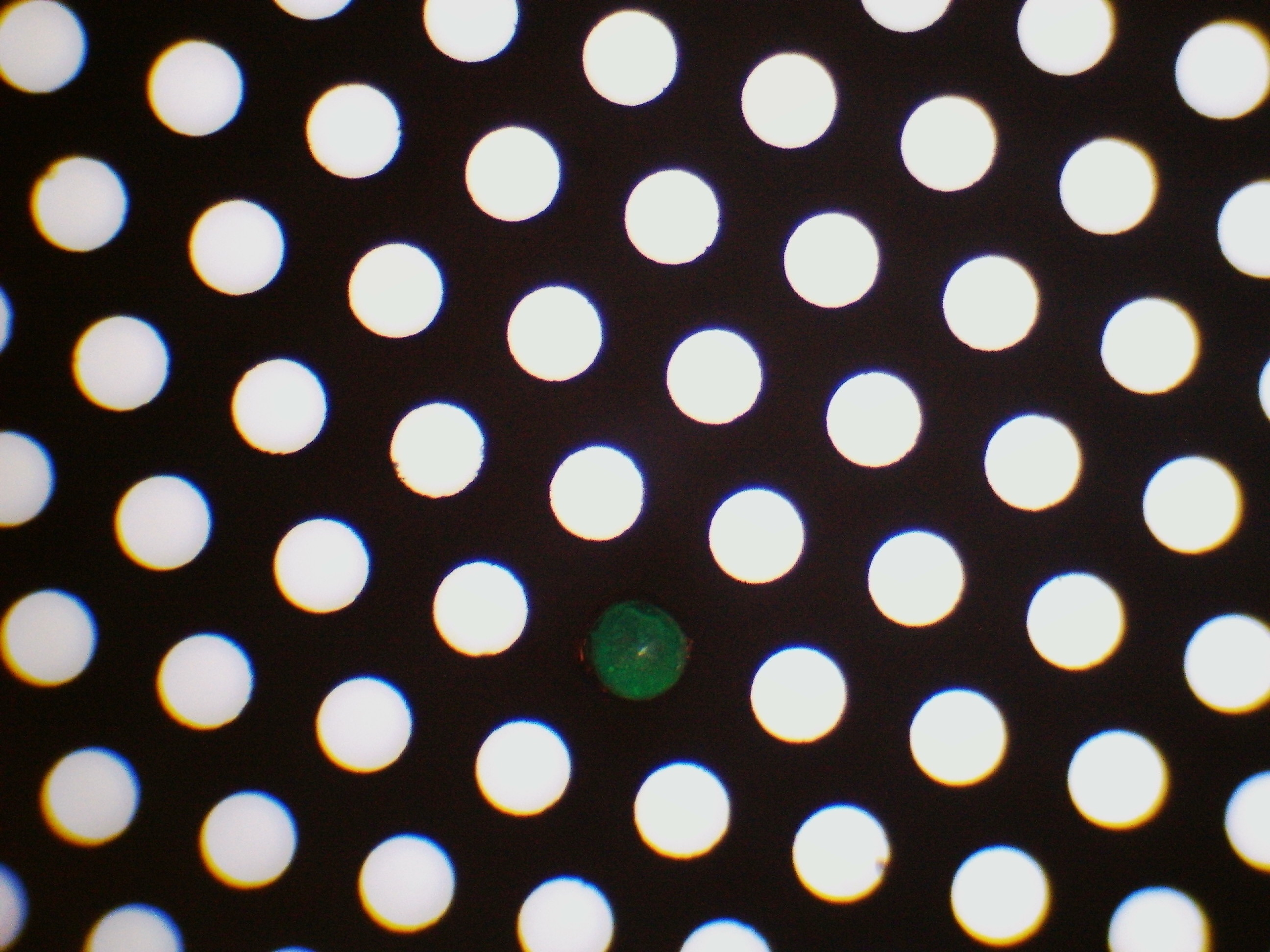}
        \caption{}
    \end{subfigure}
    \caption{(a) Diagram of experimental setup used to measure the MH-PSF from a light source in a single GEM hole. Green corresponds to light leaving GEM toward the camera, Red corresponds to light traveling through GEM substrate. (b) Microscopic view isolating a single GEM hole with tape. (c) Microscopic view showing phosphorescent paint constrained to a single GEM hole.}
    \label{fig:MethodsPhase1}
\end{figure}

\begin{figure}[pos=htbp]
    \centering
    \begin{subfigure}[t]{0.48\linewidth}
        \centering
        \includegraphics[width=1\textwidth]{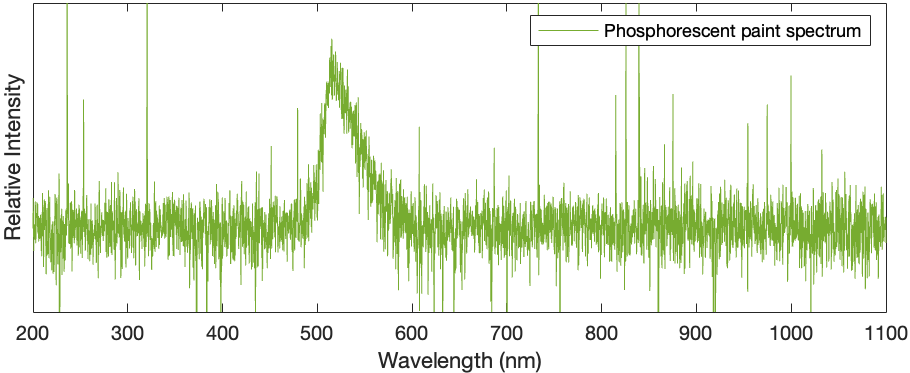}
        \caption{}
    \end{subfigure}
    \begin{subfigure}[t]{0.48\linewidth}
        \centering
        \includegraphics[width=1\textwidth]{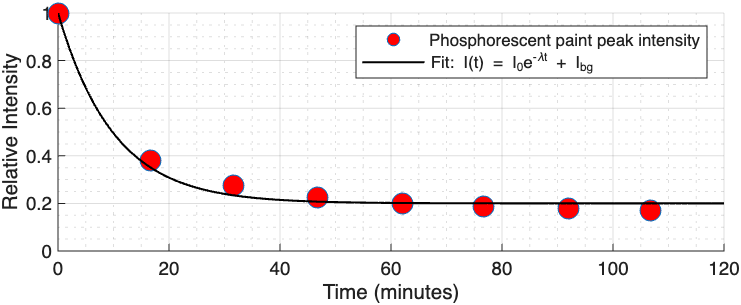}
        \caption{}
    \end{subfigure}
    \caption{Phosphorescent paint characteristics used for the MH-PSF measurements. (a) Emission spectrum showing a peak near \(\SI{520}{nm}\). (b) Time dependence of the paint intensity after excitation.}
    \label{fig:MethodsPhase1-2}
\end{figure}

We isolated a hole by covering the surrounding region with tape and, using a \SI{0.1}{\uL} pipette, inserted a 1:1 mixture of water and phosphorescent green glow-in-the-dark paint into the hole\footnote{Magicfly glow-in-the-dark green paint.\newline https://www.imagicfly.com/products/glow-in-the-dark-paint}. This dilution was selected as a practical compromise: it lowered the paint viscosity enough for repeatable pipette placement while maintaining sufficient brightness for imaging. The concentration was not systematically optimized, and variations in the final paint distribution inside the hole are treated as a limitation of the method. A diagram of the experiment is shown in Figure \ref{fig:MethodsPhase1}a and a photo of the taped hole in Figure \ref{fig:MethodsPhase1}b. After allowing time for the paint mixture to dry, the tape was removed and the area was microscopically inspected to ensure that the paint filled the hole without extending beyond either surface of the GEM. Figure \ref{fig:MethodsPhase1}c shows a single microscopic image verifying the paint is confined to a single hole; numerous microscopic images of each GEM with varied angles and lighting were used to confirm this. The G-GEM was further inspected with 3D measurements of both front/back surfaces using a Keyence VR-3000 G2 measurement system, confirming the paint does not extend beyond either surface of the GEM.
Figure \ref{fig:MethodsPhase1-2}a provides the spectrum of the green glow after exposure to the UV source for five minutes, which peaks at $\sim$520 nm\footnote{Scintillation of CF$_4$ is expected in the optical region peaking $\sim$620 nm \cite{Brunbauer-2025}, thus the paint is expected to exhibit similar optical behavior through the substrate.}, and Figure \ref{fig:MethodsPhase1-2}b shows how the light decays (I vs t) after the UV is turned off, which we roughly approximate by an exponential with a half-life of $\sim$10 minutes. 
Although repeated measurements and a full uncertainty analysis would improve the precision of the fitted decay time, this measurement was intended only to confirm that the phosphorescent emission persisted long enough for imaging and that the paint required UV re-excitation between frames.
The spectrum was recorded using an Ocean Optics HR4000CG-UV-NIR spectrometer with a detection range of 200-1100 nm, connected to fiber optic cable with a \SI{50}{\micro\meter} diameter core for light entrance. The UV source is a Wondsunson UV Light Flashlight, designed as a geological phosphor detector using 365 nm wavelength emission for phosphor excitation.

The procedure described above worked for all except the standard thin GEM, whose hole diameter and pitch were too small to insert paint into a single isolated hole. 
For the remaining 3 GEMs (G-GEM, THGEM, M-THGEM), phosphorescence images of the paint-filled holes were taken using the  Andor iKon-L 936 BV\footnote{https://andor.oxinst.com/products/ikon-large-ccd-series/ikon-l-936} CCD camera with a 2048$\times$2048 array of 13.5$\times$\SI{13.5}{\micro\meter} pixels, 16-bit digitization, 100,000 e$^-$ well depth, 2.9 e$^-$ read noise at the 0.05 MHz readout rate, and a dark current of 0.00040 e$^-$ pixel$^{-1}$ s$^{-1}$ when operating at $-70^\circ$C using a 4-stage peltier cooler. No binning was used. The camera was coupled to a Nikon AF Nikkor 50mm f/1.8D lens, providing a pixel-scale\footnote{
Pixel-scale here refers to the linear size of the GEM surface subtended by a pixel.} of \SI{9.86}{\micro\meter} per pixel. Images were acquired using the following procedure: \begin{enumerate}
    \setlength{\itemsep}{0pt}
    \item The paint-filled GEM hole was exposed to 365 nm light from the UV source for 5 minutes.
    \item The GEM was used as a passive device and, together with the camera, placed in open air. The GEM was mounted in front of the camera lens and both were draped with black cloth. The whole experiment was covered with a dark box to minimize external light. Room lights were turned off during exposures.
   \item The GEM was imaged with the Andor camera. To limit optical blooming, exposure times for each GEM were chosen so that peak pixel intensity fills 50-55$\%$ of the ADU well of the 16-bit pixel. This resulted in a 600 s per frame exposure time for the G-GEM and 30 s per frame exposure time for the remaining two GEMs.
   \item Steps 1-3 were repeated 10 times for the same paint-filled hole. This resulted in 10 exposures (science frames) used in the analysis.
   \item 
Image calibration was performed using three calibration data sets: dark frames, flat frames, and flat-dark frames. For each measurement, 10 science frames were median stacked\footnote{The median was used when stacking images to reduce the influence of pixel-intensity outliers, such as those produced by cosmic rays.}, the median stack of 10 dark frames was subtracted, and the result was divided by a master flat frame.
Dark frames were taken with the same exposure time and cooling as science frames, but with the shutter closed and lens cap on. The master flat frame was created from the median stack of 10 flat frames, then subtracting the median stack of 10 flat-dark frames.
   Flat frame images were taken with a HSK A4 artist tracing light box directly in front of the lens to provide a uniform light source, with exposure chosen with the procedure described in step 3 above.
   Flat-dark frames were taken with the same exposure time and cooling as flat frames, but with the shutter closed and the lens cap on.
\end{enumerate}

Before showing results, we note several method-specific experimental limitations associated with the optical properties of the glow-in-the-dark paint used in our studies.
The measured MH-PSF is expected to depend on emission wavelength because the GEM substrate properties (transmission, refractive index, surface reflectivity), lens transmission and camera quantum efficiency all vary across the relevant optical spectrum. Refractive-index dispersion in the GEM substrate may also change the angular distribution and total internal reflection conditions for photons propagating through the substrate, altering the relative amount of light transported to neighboring holes. In addition, the measured MH-PSF depends on the paint geometry within the hole: non-uniform filling, preferential localization near one side of the hole, or emission from the hole wall rather than the full hole volume can change both the symmetry and relative intensity of the neighboring-hole pattern. 
None of these effects change the main conclusion of these studies, that light propagates through the GEM substrate to produce multi-hole optical redistribution.

\section{\label{sec:Results}Results}

\begin{figure}[pos=htbp]
    \centering
    \begin{subfigure}[t]{0.24\textwidth}
        \centering
        \includegraphics[width=1\textwidth]{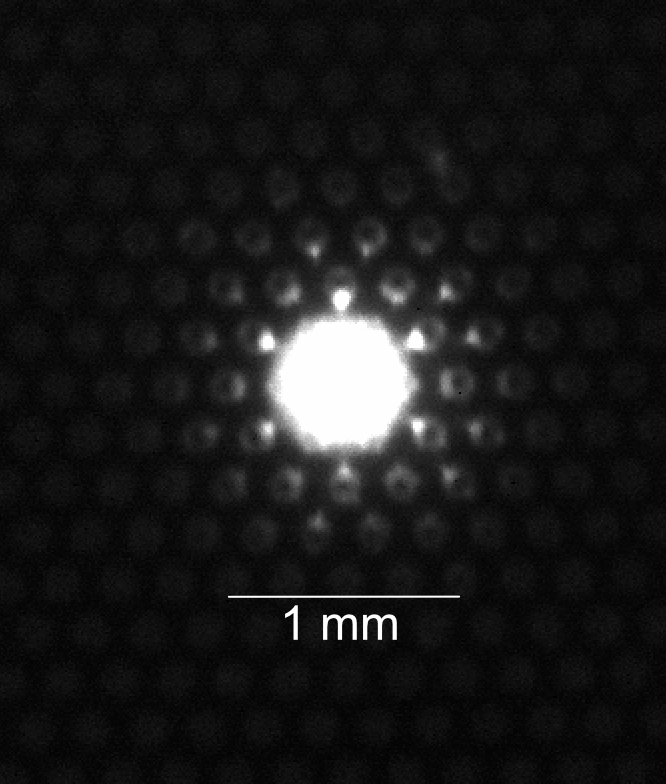}
    \end{subfigure}
    \begin{subfigure}[t]{0.36\textwidth}
        \centering
        \includegraphics[width=1\textwidth]{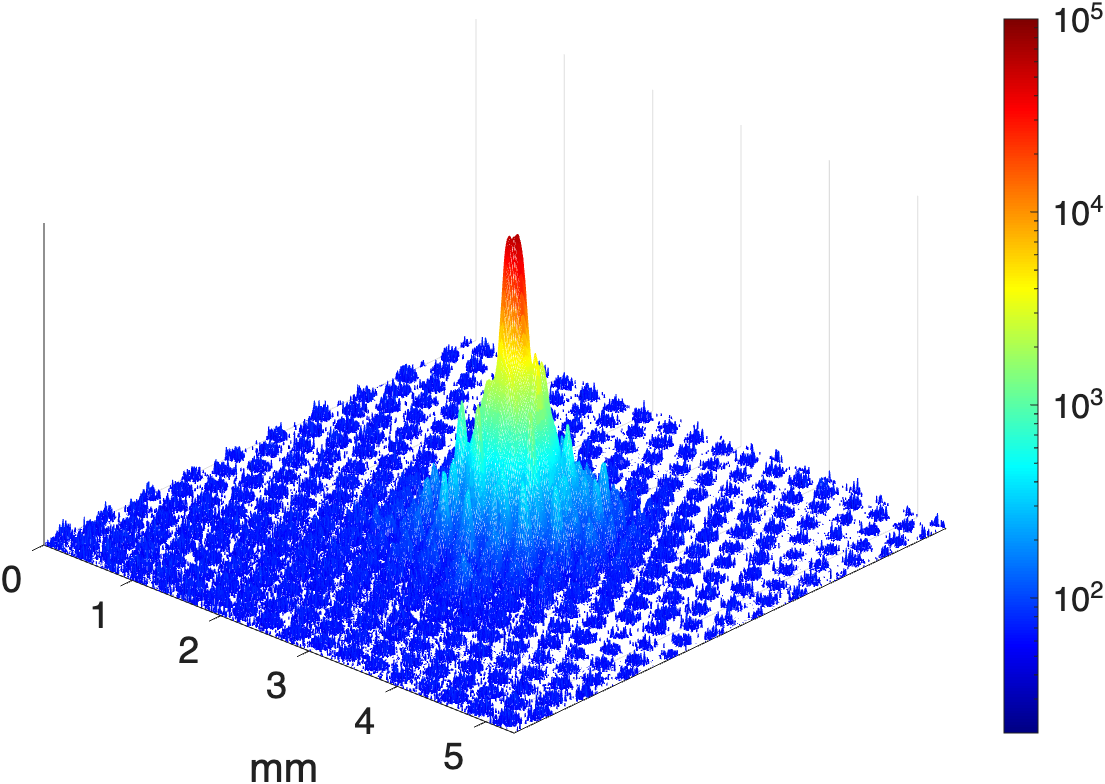}
        \captionsetup{labelformat=empty}
        \caption{G-GEM}
    \end{subfigure}
    \begin{subfigure}[t]{0.36\textwidth}
        \centering
        \includegraphics[width=1\textwidth]{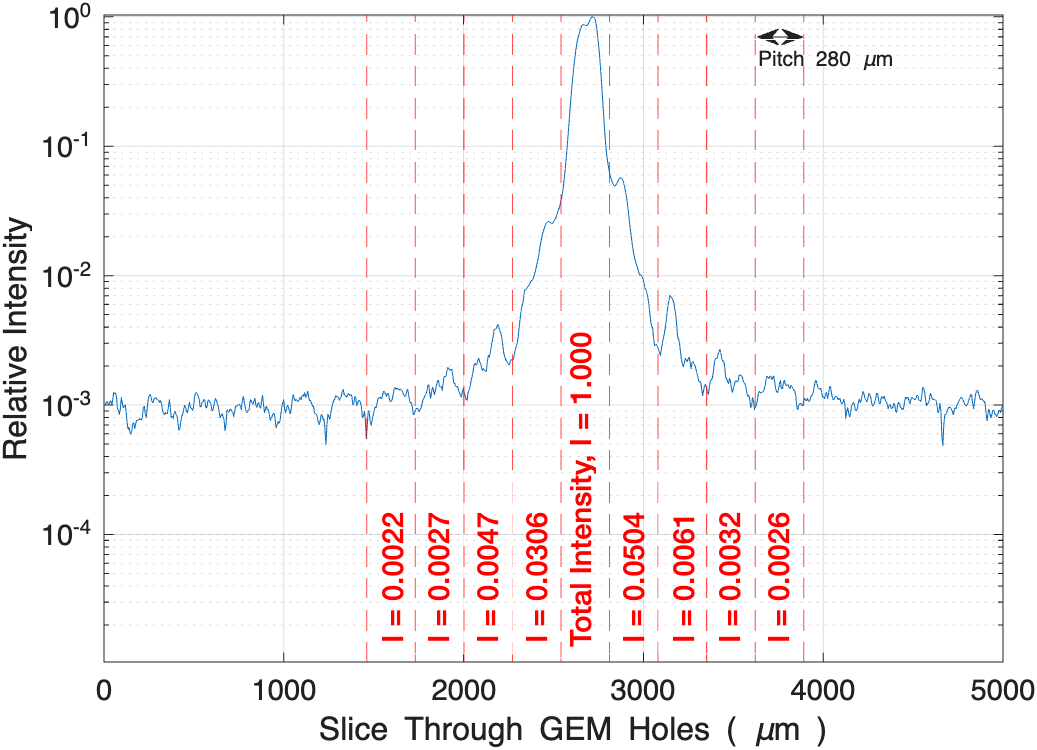}
    \end{subfigure}
    \begin{subfigure}[t]{0.24\textwidth}
        \centering
        \includegraphics[width=1\textwidth]{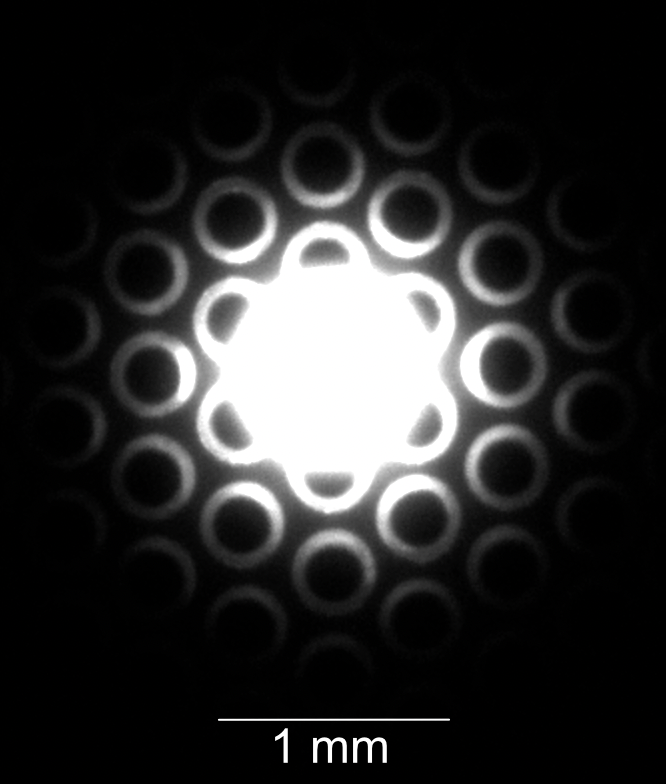}
    \end{subfigure}
    \begin{subfigure}[t]{0.36\textwidth}
        \centering
        \includegraphics[width=1\textwidth]{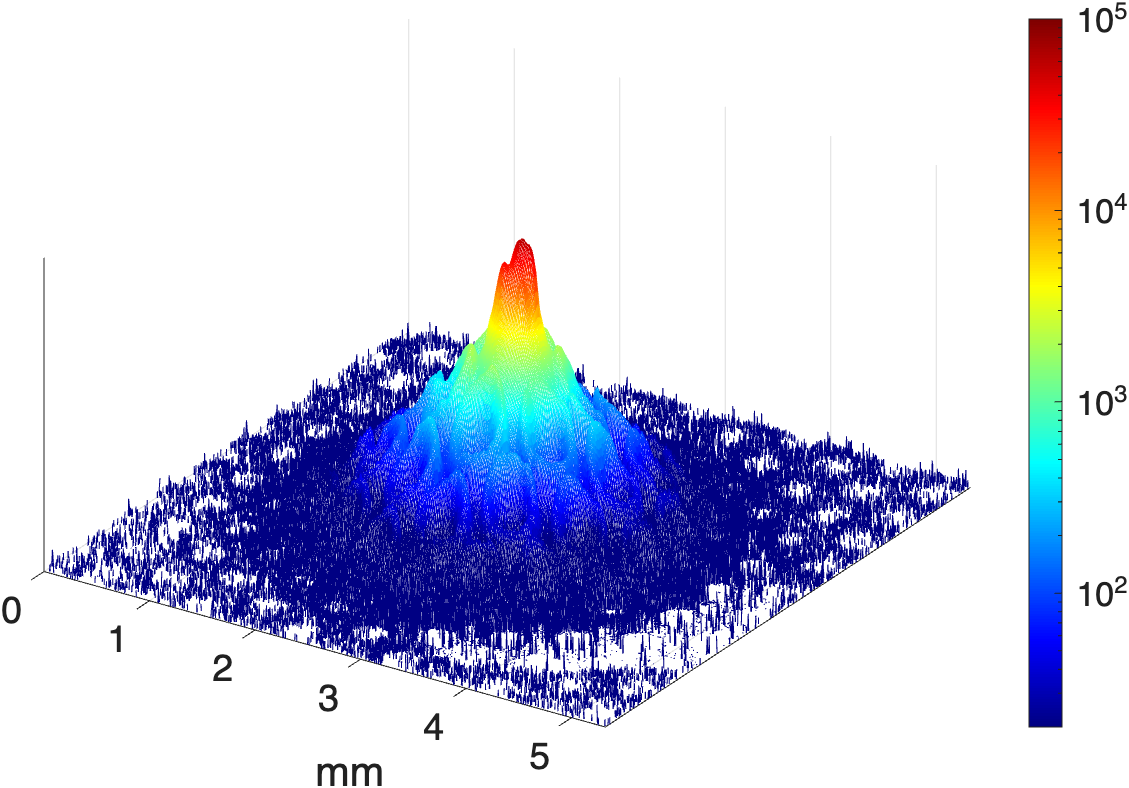}
        \captionsetup{labelformat=empty}
        \caption{THGEM}
    \end{subfigure}
    \begin{subfigure}[t]{0.36\textwidth}
        \centering
        \includegraphics[width=1\textwidth]{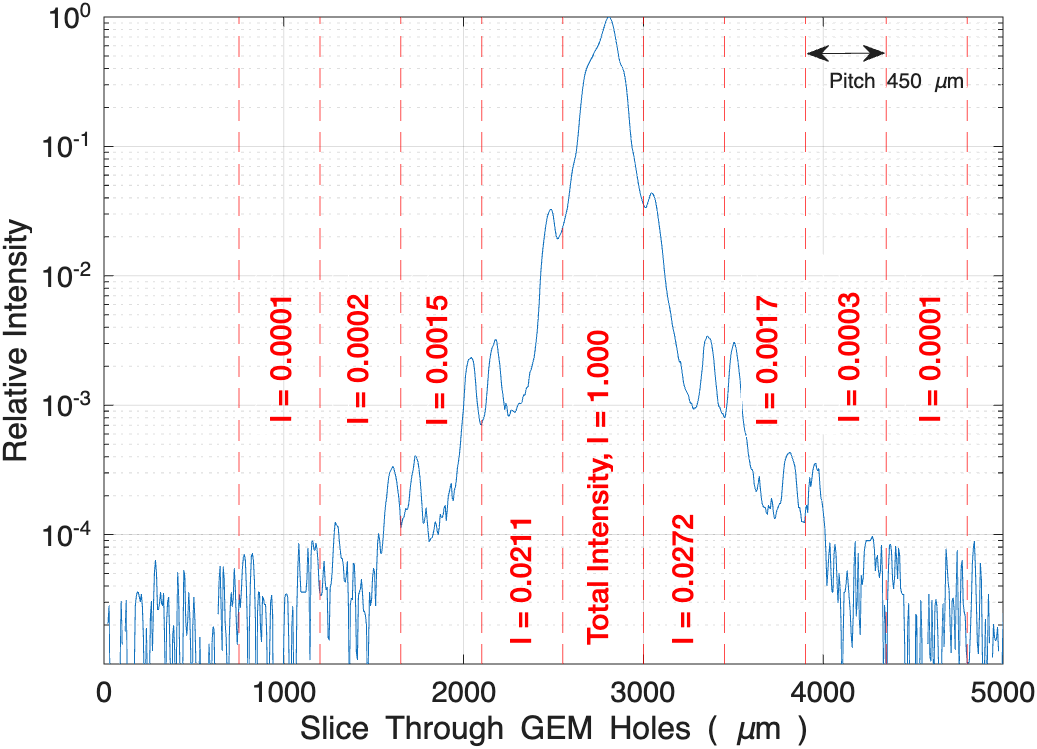}
    \end{subfigure}
        \begin{subfigure}[t]{0.24\textwidth}
        \centering
        \includegraphics[width=1\textwidth]{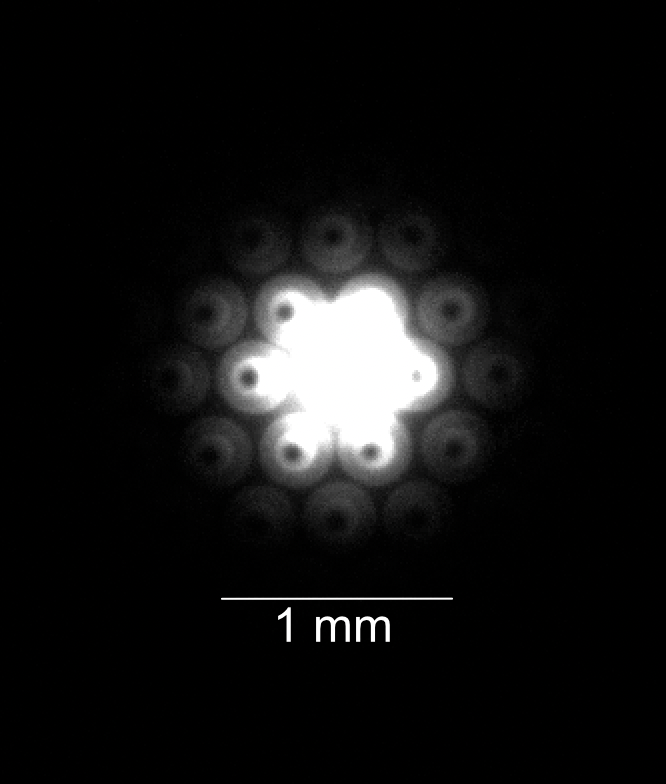}
    \end{subfigure}
        \begin{subfigure}[t]{0.36\textwidth}
        \centering
        \includegraphics[width=1\textwidth]{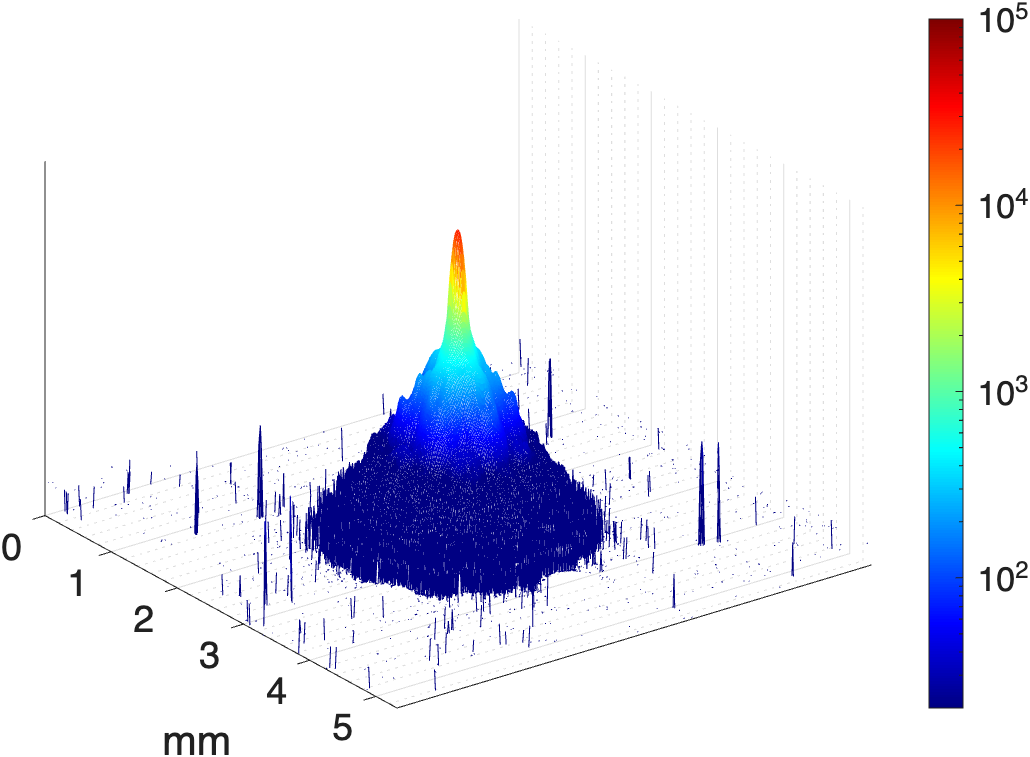}
        \captionsetup{labelformat=empty}
        \caption{M-THGEM}
    \end{subfigure}
        \begin{subfigure}[t]{0.36\textwidth}
        \centering
        \includegraphics[width=1\textwidth]{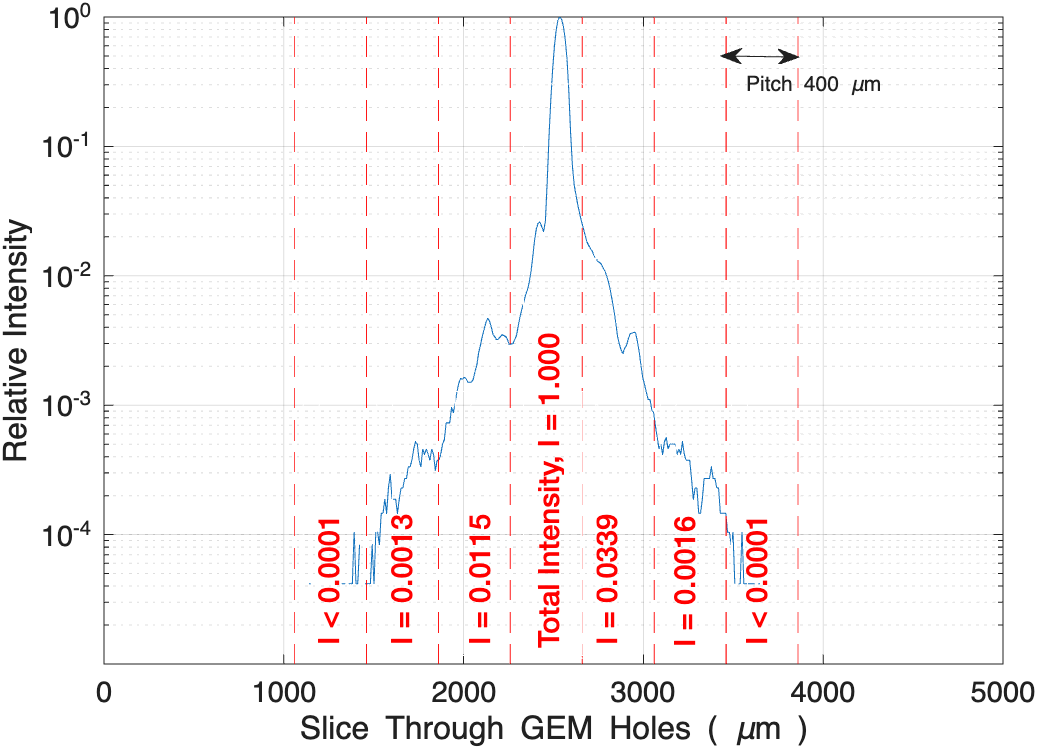}
    \end{subfigure}
    \caption{MH-PSF results from single-hole light-source measurements. Rows show the G-GEM with a 600 s exposure (top), THGEM with a 30 s exposure (middle), and M-THGEM with a 30 s exposure (bottom). Columns show the linear-scale camera image (left), logarithmic-intensity surface plot of the camera image (center), and logarithmic-intensity 1D slice through the GEM holes (right). The surface plots and 1D slices are shown on common scales within each column to allow comparison between GEM types. In the camera images, the slices pass through the peak-intensity pixel and follow the hexagonal hole-pattern orientation of each GEM: left-right for the G-GEM, top-bottom for the THGEM, and left-right for the M-THGEM.}
    \label{fig:ResultsPhase1}
\end{figure}

The results of the measurements described above demonstrate the emergence of a MH-PSF in all three GEM substrates, as shown in Figure \ref{fig:ResultsPhase1}.
The camera images in the left column clearly reveal that light produced in a single (primary) hole propagates through the GEM substrate with some fraction exiting neighboring holes. The intensity and distribution of the resulting MH-PSF vary depending on the GEM substrate and geometry (thickness, hole size, rims, etc.). The rims in the metal have a large impact on the observed light emission visible in the THGEM and M-THGEM, with the rings of light matching the visible rims seen in room-lit focusing images. The middle column shows surface plots of the 2D MH-PSFs with the intensity, normalized to that emitted by the primary hole, represented by the third dimension and with the same physical scale used for each GEM.
The right column shows intensity histograms through a slice in the hexagonal pattern that passes through neighboring holes. These are also normalized to the intensity of the primary hole. In red the total intensity for each hole is integrated along the 1D slice and shown relative to that of the primary hole. As noted in Section \ref{sec:Methods}, the M-THGEM (bottom row) demonstrates an asymmetry,
likely due to the paint being non-uniformly distributed within the hole and preferentially localized toward one side.
Each GEM demonstrates some level of asymmetry, as revealed by the slices shown in the right column. While only one slice is shown in each GEM plot to demonstrate the asymmetry, slices through the three unique rows of holes\footnote{Slices are at $0^\circ$, $60^\circ$, $120^\circ$.} in each GEM were compared and found to be consistent (after averaging the asymmetries described in Section \ref{subsec:SimulationResults}) and have been averaged to determine the MH-PSF results in Table \ref{tab:Results}. 

The uncertainties reported in Table \ref{tab:Results} are obtained by combining measurement scatter, image-noise propagation, and pixel-scale uncertainty in quadrature. The quoted uncertainties do not include possible variations associated with repeated paint placement, different holes, or different GEM samples. The dominant contribution was the measurement scatter, defined as the standard deviation of the six symmetry-equivalent measurements at each pitch distance, corresponding to the left and right neighboring holes along the three GEM axes, extracted from the same stacked image. The image-noise contribution was obtained by propagating the relative pixel uncertainties from shot noise, dark-frame noise, and flat-field correction. Because each reported value is calculated as the neighboring-hole intensity divided by the central-hole intensity from the same calibrated image, exposure time and overall image-brightness normalization enter as common multiplicative factors and cancel in the relative intensity. The pixel-scale was determined from focus images in which the GEM hole pattern was visible across the imaging plane, allowing the pitch to be measured over many hole spacings rather than from adjacent holes alone. This made the pixel-scale uncertainty small. Repeating the analysis with the pitch shifted by its estimated uncertainty did not change the selected integer-pixel for each neighboring GEM hole and therefore gave a negligible contribution to the combined error.

As might be expected, the G-GEM is more transparent than both the THGEM and M-THGEM. Conversely, the M-THGEM is the least transparent, as its substrate is ceramic. A quantitative analysis, shown both in Figure \ref{fig:ResultsPhase1} and in the third column of Table \ref{tab:Results}, shows that the relative intensities of neighboring holes in the glass GEM are $>1.5$ times that of the others and approach a nonzero pedestal at larger pitch distances. The origin of the pedestal is discussed further in Section \ref{sec:Simulations}.

\section{\label{sec:Simulations}Simulating an accurate MH-PSF}

While our experimental studies enabled the measurement, characterization, and comparison of MH-PSFs in three different GEM materials, the phosphorescent-paint measurements represent only the first step toward understanding GEM-induced optical broadening in realistic detector conditions. The paint measurements provide an experimental benchmark for validating Geant4 optical simulations of photon transport through a GEM. Once benchmarked against these measurements, the simulations can be extended to configurations that are difficult to reproduce experimentally with the paint method, including multiple stacked GEMs, detector-specific reflected-light contributions, and photon starting distributions derived from Garfield++\cite{garfieldpp} avalanche simulations.

In this section, we develop Geant4 simulations of MH-PSFs for both a single G-GEM paint model, for comparison with our measurements, and a double G-GEM configuration with photon starting distributions derived from Garfield++ avalanche simulations. Given the significant impact GEM optical effects could have in experiments, an accurate simulation of the MH-PSF for realistic experimental conditions is therefore attractive. Characterizing these effects is also a necessary step toward developing a deconvolution method with the potential to undo much of the GEM-induced broadening of tracks. The task of developing a deconvolution method is not covered in this work.

Our simulation begins with a Geant4 model of a small region of the G-GEM (Section \ref{subsec:GEMmodeling}). Geant4 is also used to generate photons inside a single hole according to a specified distribution, and track them through the GEM and other regions of interest (Section \ref{subsec:OpticalSim}). An approximate method for image formation that results in the MH-PSF is detailed in Section \ref{subsec:PhotonPosition}. The full simulation described here is validated against our glowing paint measurements in Section \ref{subsec:SimulationResults}, and then used to predict the MH-PSF expected in realistic experiments in Section \ref{subsec:SimulationMIGDAL}.

\subsection{\label{subsec:GEMmodeling}GEM Modeling}

\begin{figure}[pos=htbp]
    \centering
    \begin{subfigure}[t]{0.22\linewidth}
        \centering
        \includegraphics[width=1\textwidth]{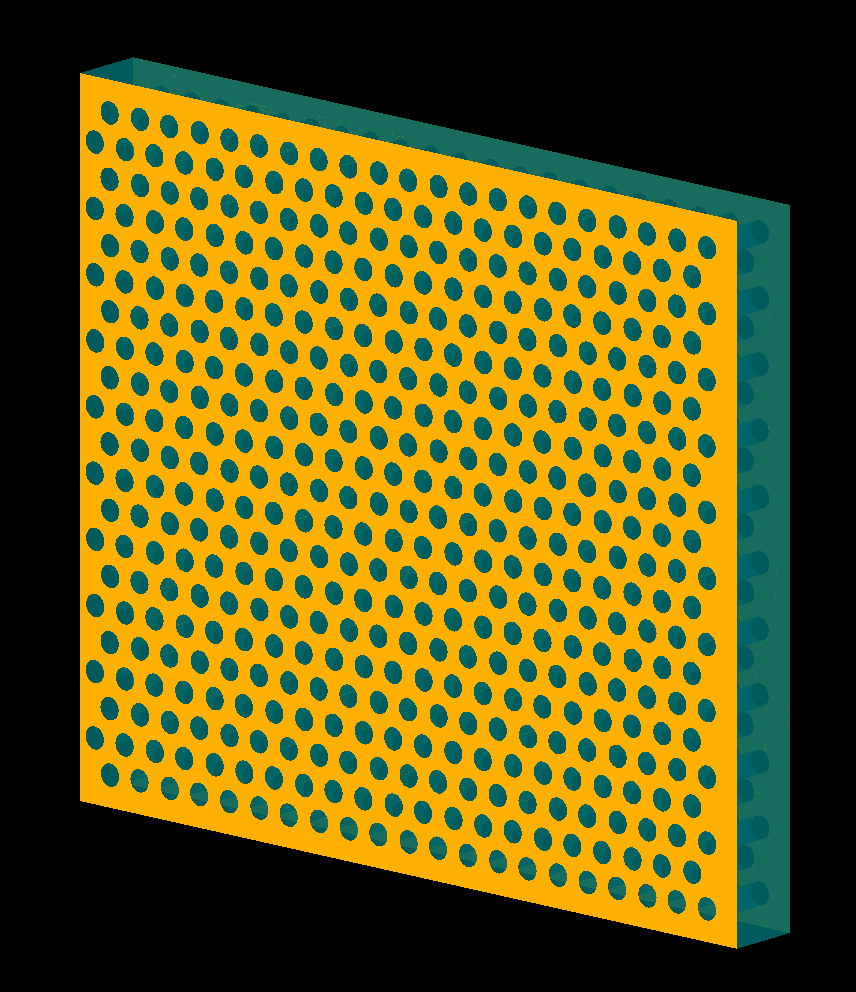}
        \caption{}
    \end{subfigure}
    \begin{subfigure}[t]{0.265\linewidth}
        \centering
        \includegraphics[width=1\textwidth]{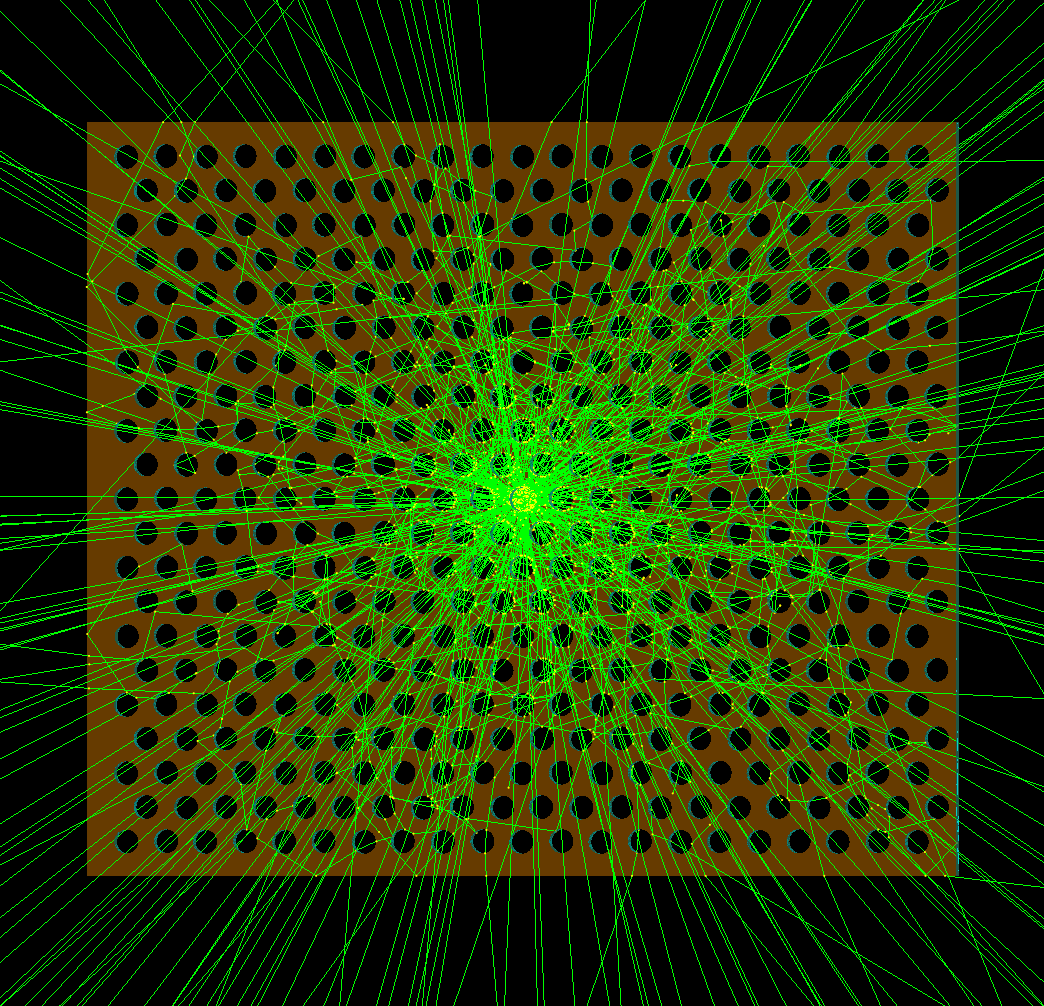}
        \caption{}
    \end{subfigure}
    \caption{Geant4 GEM simulation geometry and photon-tracking example. (a) A \(21\times21\)-pitch section of the simulated G-GEM, with dimensions matching those listed in Table~\ref{tab:Results}. (b) Example optical-photon trajectories for 500 photons generated inside a single G-GEM hole, illustrating photon propagation and interactions within the GEM structure.}
    \label{fig:Sim_1}
\end{figure}

GEMs can be modeled in Geant4 by creating separate layers for the substrate and electrodes with physically accurate thicknesses. We chose a small area of a G-GEM to simulate,
11.76$\times$11.76 mm, which provides a grid of 41$\times$41 GEM holes (20 holes on each side of the central hole containing the light source) sitting within a world volume vacuum of a cubic meter. Chosen to save on computational expense, the truncated area is still sufficiently large to accurately model the MH-PSF.
A smaller portion (21$\times$21 GEM holes) of the full geometry (41$\times$41 GEM holes) used in simulation is shown in Figure \ref{fig:Sim_1}.

\subsection{\label{subsec:OpticalSim}Optical Photon Simulation}

The Geant4 simulations were performed as optical-transport simulations in vacuum, with photons generated directly inside the GEM geometry rather than produced through a gas-avalanche or scintillation process. Optical-photon simulations begin by generating particles of the Geant4 particle type \texttt{opticalphoton} with a particle gun, emitted isotropically with a fixed energy of $\sim$\SI{2}{eV}.\footnote{Scintillation of CF$_4$ is expected to peak $\sim$620 nm in the optical region \cite{Brunbauer-2025}, which corresponds to a photon energy of $\sim$\SI{2}{eV}.}
Photon starting locations can be fixed at the center of a GEM hole or sampled from a specified spatial distribution. Section \ref{subsec:SimulationResults} uses point-source and cylindrical paint distributions, while Section \ref{subsec:SimulationMIGDAL} uses a distribution derived from Garfield++ avalanche simulations.
\texttt{G4OpticalPhysics} \cite{Allison_2016} (or another appropriate optical physics list) must be added to the physics list and any energy limits used must include $\sim$\SI{2}{eV} photons. Lastly, the geometry must be modified to include boundary properties for the interface of materials as well as the optical properties of each material. Boundary properties include the type (dielectric-to-dielectric, dielectric-to-metal, etc.), reflectivity/transmittance, and surface finish. Optical properties for a material include the refractive index and absorption length. 

Our model of the G-GEM 
utilizes the following boundary properties: glass/copper surface set as dielectric-to-metal; vacuum/glass surface set to dielectric-to-dielectric; and vacuum/copper surface set as dielectric-to-metal. All surfaces are polished. All reflections are computed in Geant4 using the optical properties for each material. The refractive indices used were $n_\mathrm{vacuum}=1$, $n_\mathrm{glass}=1.535$ \cite{PEG3refractive}, and a complex refractive index for copper of $n_\mathrm{copper}=0.27111$ and $k_\mathrm{copper}=3.2466$ \cite{CopperRefractive}. The absorption length for glass is negligible in optical wavelengths \cite{PEG3refractive} so is set to \SI{10}{m}, and for copper it is set to \SI{15}{nm} \cite{CopperRefractive}.

The optical simulations use simplified material and boundary assumptions intended to isolate the MH-PSF mechanism, rather than to provide a complete detector-specific optical model. The resulting MH-PSF should therefore be interpreted for the specific optical assumptions used here. Changes in detector-specific material properties, surface conditions, reflectivity, or imaging-system response could modify the quantitative MH-PSF, and evaluating those dependencies would require a dedicated optical characterization for a given detector. Such a detector-specific study is beyond the scope of this work.

\subsection{\label{subsec:PhotonPosition}Image Formation}

The Geant4 simulation tracks every photon in the GEM until it leaves the GEM surfaces. The position and direction of every photon that exits the surface facing a lens-camera is recorded. An approximate method that avoids the complexity of full ray-tracing required to model the effects of real lens-camera optical systems is used. The known photon position and direction allow a disk intersection method of determining if a given photon would be recorded by a camera. This method captures the dominant geometrical acceptance needed to estimate the MH-PSF in different experimental configurations, while intentionally neglecting detailed lens aberrations, finite depth of field, focus errors, diffraction, and wavelength-dependent imaging effects.

The Geant4 simulations generate $10^{9}$ photons inside a G-GEM hole using different distributions, described in Sections \ref{subsec:SimulationResults} and \ref{subsec:SimulationMIGDAL}. The photons are followed until they leave the world volume or they enter a ``detector" surface used to form the image. This detector of material \texttt{G4\_Galactic} (to represent low density vacuum) is placed directly against the surface of the G-GEM. For each step taken by a photon, Geant4 checks if it is within the detector volume and if it is traveling toward the direction a camera lens would be placed. If both are true, the photon position and direction are recorded and the photon is terminated to begin the next simulated photon. 
The Geant4 direction check before recording is needed to allow photons to be created just beyond the GEM hole, as is found in realistic charge avalanche simulations. Photons created here may travel toward the GEM and add to the MH-PSF or toward a camera, so they are treated separately.

\begin{figure}[pos=htbp]
    \centering
    \begin{subfigure}[t]{0.49\textwidth}
        \centering
        \includegraphics[width=1\textwidth]{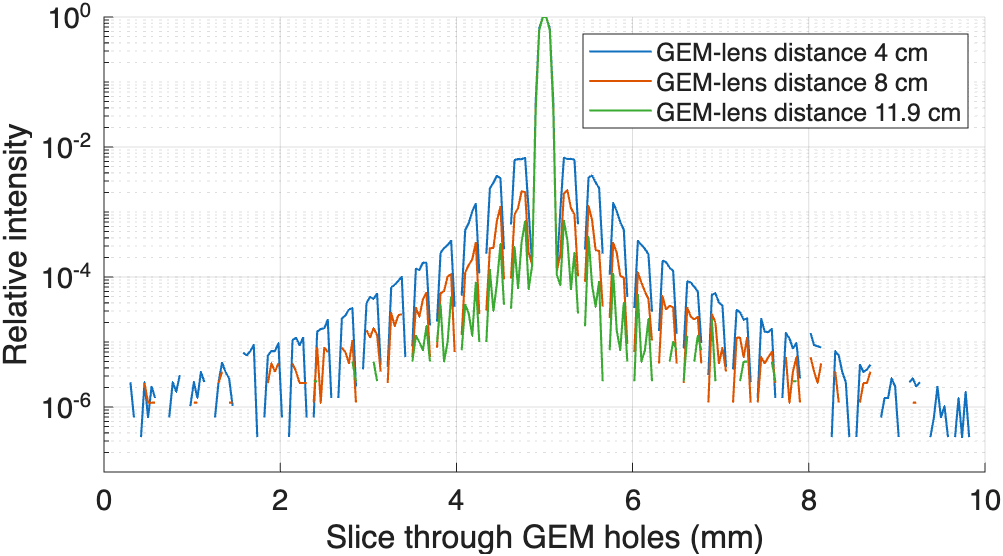}
        \caption{}
    \end{subfigure}
    \begin{subfigure}[t]{0.49\textwidth}
        \centering
        \includegraphics[width=1\textwidth]{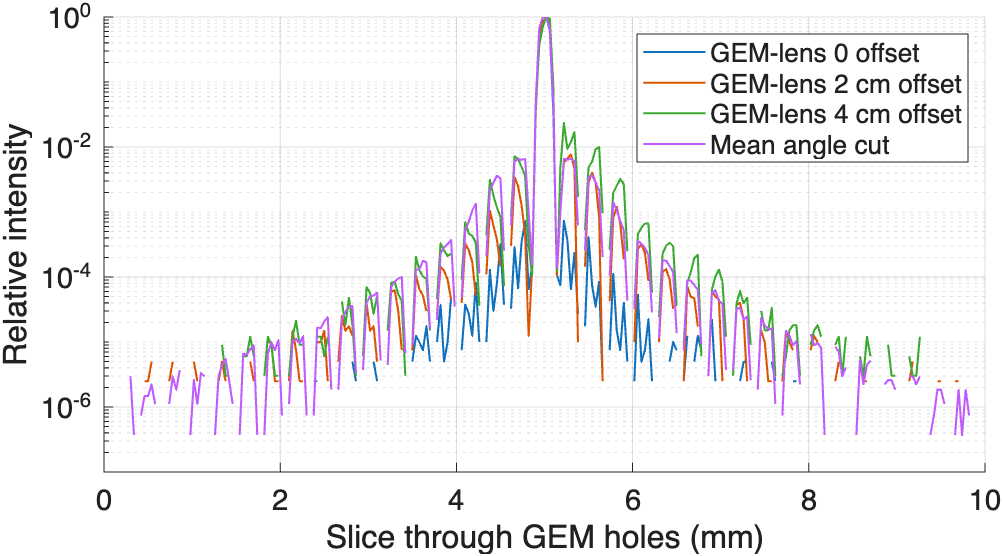}
        \caption{}
    \end{subfigure}
    \caption{Geant4 MH-PSF sensitivity to lens position for the G-GEM simulation. One-dimensional slices through the simulated GEM-hole pattern show how the MH-PSF changes with (a) GEM-to-lens distance and (b) transverse lens offset, perpendicular to the GEM-to-lens axis. The slices compare the redistribution of photon intensity among neighboring GEM holes for each lens configuration. The sharp downward excursions arise from zero simulated intensity in the copper regions between GEM holes.}
    \label{fig:SimOffset}
\end{figure}

Using a known lens distance and diameter, a Python script can read the Geant4 output and calculate if each photon would hit or miss a camera lens, placed at a distance relative to the GEM, matching any given experiment.
For each photon that hits the lens, the initial position at the bottom of the GEM is used under the assumption that the lens would focus light back to that point (the focus of the camera system). This introduces a dependence of the relative positions of the GEM hole emitting light and camera lens. 
Figure \ref{fig:SimOffset}a reveals the MH-PSF decreases as lens distance increases. The plot contains sharp downward excursions between GEM holes because the simulated intensity is zero in the copper regions, where no photons are emitted. Figure \ref{fig:SimOffset}b shows the change to the MH-PSF as the center of the camera lens is shifted relative to the GEM center (shifted perpendicular to GEM-lens distance). 
When the GEM and lens centers are perfectly aligned (blue), the MH-PSF has the least intensity from neighboring holes. As the lens center is offset \SI{2}{cm} (red) and \SI{4}{cm} (green), the intensity from neighboring holes grows with distance, preferentially from holes closer to the lens center (right side of Figure \ref{fig:SimOffset}b).

\begin{figure}[pos=htbp]
    \centering
    \begin{subfigure}[t]{0.48\linewidth}
        \centering
        \includegraphics[width=1\textwidth]{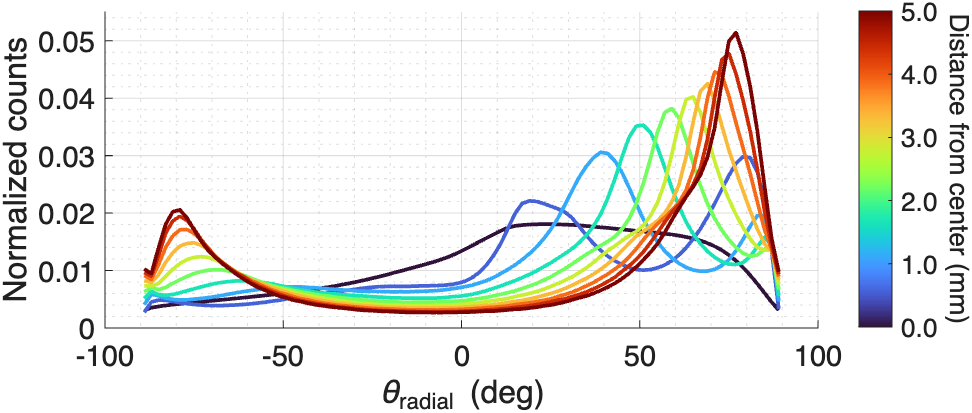}
        \caption{}
    \end{subfigure}
    \begin{subfigure}[t]{0.48\linewidth}
        \centering
        \includegraphics[width=1\textwidth]{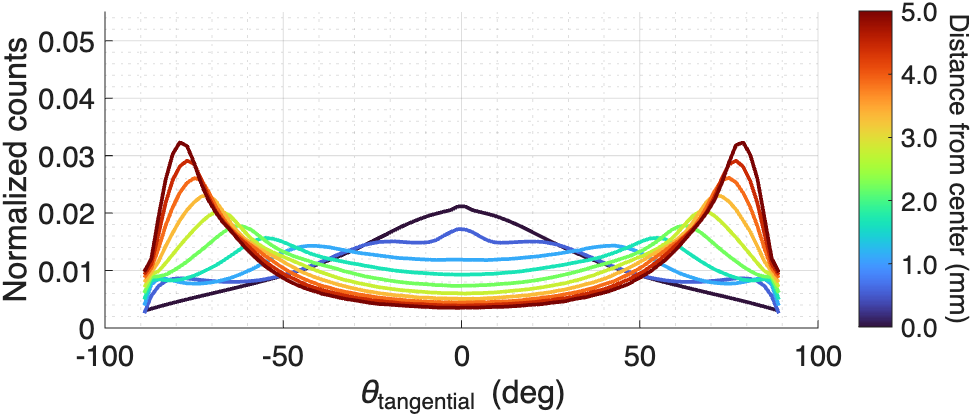}
        \caption{}
    \end{subfigure}
    \caption{Angular distributions of photons exiting the G-GEM holes after propagation through the substrate. The x-axis gives the photon exit angle relative to the surface normal, where \(0^\circ\) corresponds to photons exiting perpendicular to the GEM surface. The y-axis shows the normalized photon counts, and the color scale indicates the distance traveled through the substrate before exiting. Distributions are shown for (a) the radial plane, corresponding to inward/outward propagation relative to the central source hole where the photons were generated, and (b) the tangential plane, perpendicular to the radial direction from the central source hole.}
    \label{fig:SimAngles}
\end{figure}

The Geant4 output direction is vital to exclude photons traveling at large $\theta$, the angle with respect to the lens direction. Figure \ref{fig:SimAngles} reveals how the photon angular distribution changes with distance from the central hole, where the light was initially emitted. $\theta$ is measured in both the radial plane and tangential plane (relative to a central hole), which show that photons emitted at greater distances from the central hole tend to leave the GEM at larger angles. Importantly, while photons are most likely to be pointed away from the central hole, radially outward (right side of Figure \ref{fig:SimAngles}a), they are also likely to be emitted at a large angle toward the central hole (left side of Figure \ref{fig:SimAngles}a) or perpendicular to it (left and right of Figure \ref{fig:SimAngles}b). These large $\theta$ scatters explain Figure \ref{fig:SimOffset}b, where the GEM-lens offset favors larger $\theta$, and more of the photons from neighboring GEM holes will be observed. The large $\theta$ with distance also explains Figure \ref{fig:SimOffset}a, as a closer lens increases the accepted angles photons could travel and still hit the lens.

The Geant4 output
allows a per-pixel MH-PSF to be created to fully account for angular effects from the entire camera field of view. However, a simplified approximation can be used, wherein an average MH-PSF across the imaging plane is determined by the geometry. 
Matching the geometry used in the MIGDAL experiment, we use a lens distance of \SI{119}{mm} and radius of \SI{14.7}{mm} to compute the maximum angular acceptance from each point on the GEM to reach the lens.
Using the known camera field of view, the maximum angle for each pixel covering an area on the GEM is calculated and a mean angle of 19$^{\circ}$ is found. Figure \ref{fig:SimOffset}b shows a slice through the MH-PSF GEM holes derived from cutting all photons greater than 19$^{\circ}$ (purple) as they are unlikely to reach the camera. This reduces the number of photons a camera would detect to $1.7\%$ of the $10^9$ simulated for the cylinder of paint distribution described in Section \ref{subsec:SimulationResults}. 

The focusing method described here is an idealized image-formation approximation. The simulated MH-PSF should be interpreted as the GEM-induced component of the optical response under idealized focusing conditions, rather than as a complete model of the final camera image. In a real camera-lens system, finite depth of field, defocus, aberrations, diffraction, sensor sampling, and imperfect alignment would broaden the image of each emitting region beyond the geometrical mapping used here. These effects would contribute an additional ordinary optical point-spread function on top of the GEM-induced MH-PSF. Chromatic dispersion is also neglected because the simulations are performed at a single photon energy. For a broadband scintillation spectrum, wavelength-dependent refractive indices, substrate transmission, lens chromatic aberration, and camera quantum efficiency could change how different wavelengths contribute to the recorded image, affecting both the amount of light transported through the GEM and the degree of optical blur.

The present simulation does not attempt to model the full lens-camera PSF. Instead, it isolates the redistribution of light caused by propagation through the GEM and applies a geometrical lens-acceptance cut to determine which photons could reach the camera. Figure \ref{fig:SimOffset} is therefore used as a sensitivity study showing how the MH-PSF depends on lens distance and transverse lens offset. The image-formation approximation is intended to capture the general behavior of GEM-induced light redistribution without assigning a detector-specific lens-alignment uncertainty. Quantifying such an uncertainty would require repeating the MH-PSF extraction over a defined range of lens-distance and offset tolerances for a specific detector configuration. The full detector-scale optical simulations discussed in Section~\ref{sec:Summary} would provide the appropriate framework for including the full GEM area, full detector reflections, lens-distance tolerances, and alignment uncertainties.

\subsection{\label{subsec:SimulationResults}Reproducing Measured MH-PSF}

\begin{figure}[pos=htbp]
    \centering
    \begin{subfigure}[t]{0.49\textwidth}
        \centering
        \includegraphics[width=1\textwidth]{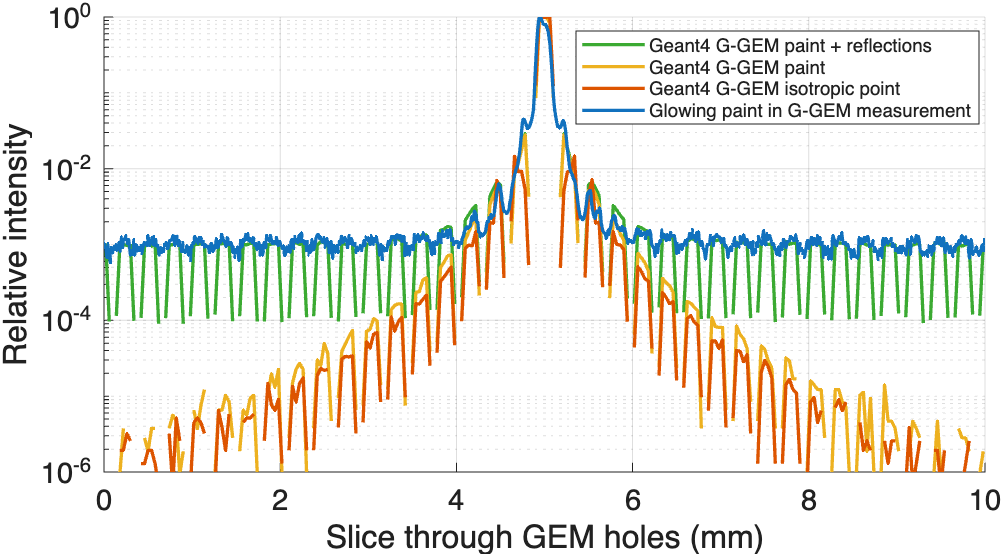}
        \caption{}
    \end{subfigure}
    \begin{subfigure}[t]{0.49\textwidth}
        \centering
        \includegraphics[width=1\textwidth]{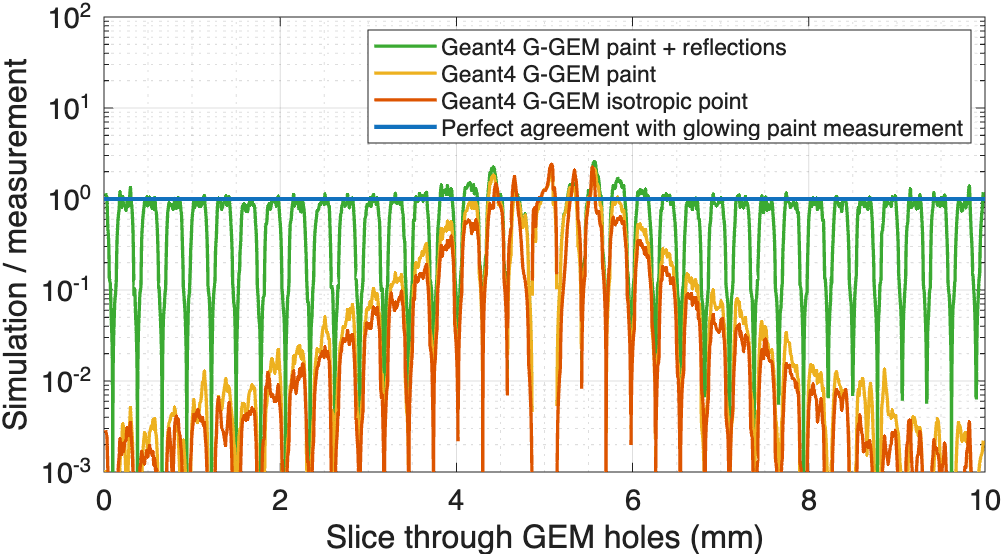}
        \caption{}
    \end{subfigure}
    \caption{(a) MH-PSF slices comparing Geant4 simulations with the glowing-paint measurement in log scale. (b) Simulation-to-measurement ratio for the MH-PSF slices in log scale. A value of \(R_i = S_i/M_i\)=1 indicates agreement with the glowing-paint measurement. The reflection-enhanced simulation remains closest to unity over the measured profile, while the isotropic point and cylinder of paint simulations underpredict the low-intensity pedestal in the outer regions. The sharp downward excursions in (a) and (b) arise from zero simulated intensity in the copper regions between GEM holes.}
    \label{fig:Sim_2}
\end{figure}

Figure \ref{fig:Sim_2}a shows the measured (for the glowing paint, shown in blue) 
and three simulated MH-PSFs for the G-GEM. The latter are for different source photon distributions, described below. The intensity plotted in the figure is 
the mean over the three unique lines of symmetry\footnote{In Figure \ref{fig:ResultsPhase1} top-left these slices are at $0^\circ$, $60^\circ$, $120^\circ$.} that pass through 
GEM holes including the central hole containing the paint (Figure \ref{fig:ResultsPhase1} only shows one slice).
This was done to average out the asymmetries and statistical fluctuations in the measured and simulated MH-PSFs, respectively.

The isotropic point source simulates photons located at the center of a G-GEM hole (red). The two paint simulations (green and yellow) attempt a photon distribution that better represents the cylindrical geometry of the paint in the hole. This is done with
a single parameter that is varied to tune 
where emission occurs within the GEM hole: the fraction of the \SI{570}{\micro\meter} hole length filled with paint (50$\%$ fill shown).
The photons are then emitted from random points on the surface of the cylinder, in a direction following a cosine-weighted Lambertian distribution 
from the surface; the Lambertian distribution is chosen to imitate surface emission.
We additionally note the possibility that, in our experimental setup, light could escape from the back side of the GEM (away from the camera), where a metal GEM mounting bracket and black cloth lie a few millimeters away. This light could reflect back towards the GEM, potentially explaining the diffuse contribution seen in our observations (the nonzero pedestal in Section \ref{sec:Results}). 
We therefore simulate an additional uniform light source across the back of the G-GEM which is shown in green in Figure \ref{fig:Sim_2}a. 

The reflected-light contribution used in this comparison is a phenomenological adjustment, not an independently measured detector reflectivity. It was included to test whether a plausible reflected component could reproduce the measured neighboring-hole intensities. Thus, the measured MH-PSF is the experimentally constrained result, while the reflection-enhanced simulation is a model-dependent interpretation of the observed pattern. The intensity for this light source was adjusted to match measurements and represents 150$\%$ of the intensity from the cylinder of paint.
While this number seems high for a background source, the intensity is integrated over the entire simulated GEM surface 
and only represents $2.6\%$ of the $10^9$ simulated photons. This result appears reasonable for the number of photons that exit the back of the GEM and reflect off surfaces back to the GEM.
This diffuse glow can be observed at a lower level in
the THGEM where the rims appear to glow away from the central hole. It is not seen with the ceramic M-THGEM. We assume that diffuse light across the back is mostly absorbed by the ceramic material and inner layers before reaching the front that is facing the camera.

To quantify the comparison between the measured and simulated MH-PSF profiles, we use the simulation-to-measurement ratio, \(R_i = S_i/M_i\), where \(S_i\) and \(M_i\) are the simulated and measured relative intensities at the same position. Shown in Figure \ref{fig:Sim_2}b,  the ratio provides a direct comparison across the full profile, including the low-intensity pedestal visible on the logarithmic intensity scale. The reflection-enhanced simulation remains closest to \(R_i = 1\) over the measured profile, while the isotropic and paint-source simulations underpredict the measured pedestal in the outer regions. For comparison metrics, the RMS fractional difference, \(\sqrt{\langle[(S_i-M_i)/M_i]^2\rangle}\), decreases from 0.92 for the isotropic simulation and 0.90 for the paint-source simulation to 0.59 for the reflection-enhanced simulation. Similarly, the RMS log-ratio, \(\sqrt{\langle[\log_{10}(S_i/M_i)]^2\rangle}\) decreases from 2.51 and 2.30 for the isotropic and paint-source simulations, respectively, to 0.83 when reflections are included, corresponding to a reduction in the typical multiplicative disagreement from factors of approximately 325 and 200 to approximately 6.7. This quantitative comparison supports the conclusion that the reflected-light component is needed to reproduce the measured low-intensity pedestal. The remaining disagreement is expected in part because the simulation produces an idealized GEM-hole profile with no photon emission from the copper regions between holes, whereas the glowing-paint measurement includes ordinary optical broadening from the imaging system and nonzero intensity between GEM holes.

From Figure \ref{fig:Sim_2} it is clear that the MH-PSF is dependent on the distribution of light produced within the G-GEM hole. The isotropic point source differs from the paint models.
The agreement between simulation and measurement shows that Geant4 can reproduce the observed single-hole redistribution pattern using physically plausible source-geometry and reflected-light contributions. However, because the paint fill fraction and reflected-light contribution are tuned phenomenological parameters rather than independently measured quantities, this comparison should be interpreted as a validation of the optical-transport mechanism rather than as a unique or fully constrained detector model.
This validation provides confidence in the use of Geant4 to simulate MH-PSFs for realistic scintillation distributions in GEM-based OTPCs.

\subsection{\label{subsec:SimulationMIGDAL}Realistic MIGDAL MH-PSF}

Both panels of Figure \ref{fig:SimOffset} show a G-GEM MH-PSF with a photon distribution derived from Garfield++ simulations of electron production in charge avalanche. Here we assume that the distribution of scintillation photons' start locations closely follows that of the electron production. 
Simulations were also created using the same model for two stacked G-GEMs as used in MIGDAL, with light produced in each G-GEM. The Garfield++ avalanche simulations, performed for detector gas and field conditions, were used only to define the spatial distribution of photon starting positions inside the GEM holes. The subsequent Geant4 simulation transported these photons in vacuum through the GEM geometry and did not simulate gas-dependent scintillation production, absorption, or scattering. MIGDAL is used here as a representative GEM-based OTPC application in which GEM-induced optical effects may affect the reconstruction of the nuclear recoil (NR) and electron recoil (ER) tracks emerging from a common vertex. The primary focus of this work is not a full MIGDAL event-reconstruction study, but the optical response of GEM structures, including light propagation through the GEM substrate, redistribution of light between neighboring holes, and reflected-light contributions to the observed camera image. These simulations are used in Section \ref{sec:Discussion}.

Section \ref{subsec:SimulationResults} describes the process of adding intensity to the MH-PSF to account for photons leaving a GEM surface and reflecting off other surfaces back onto the GEM. A relatively large reflected-light contribution was required to reproduce the paint measurements because reflective surfaces were located directly behind the GEM. MIGDAL uses two stacked G-GEMs separated by \SI{2}{mm}, each with reflective copper layers. The GEMs are operated inside a reflective aluminum vacuum chamber. Geant4 predicts $<3.2\%$ of photons generated in two stacked G-GEM charge avalanche will be detected by the MIGDAL lens parameters. With $\sim97\%$ of photons bouncing around inside the vacuum chamber, realistic MH-PSFs will have to account for the expected reflections for a given experiment. 

Detailed simulations of the expected reflections in MIGDAL are left for future work, and in this work we take the same approach from recreating paint measurements and tune the intensity added to match observations. In Section \ref{sec:Discussion} we use a minimum modeled MH-PSF that shows the change to particle tracks from the GEM structure alone, without any reflective surfaces outside the GEM. This represents photons that travel directly from GEM to the lens, including any GEM structure optical redistribution. This represents the lower-broadening scenario considered in this study and is not intended to constitute a general physical lower bound. We also apply a MH-PSF with an additional uniform light source across the back of the G-GEM to account for reflections. Reflected light is treated as a detector-dependent modeling assumption rather than a calibrated optical property. Because the relevant surface reflectivities, roughness, wavelength dependence, and diffuse-reflection components were not independently measured, the reflection-enhanced MH-PSF should be interpreted as a model-dependent estimate of the possible optical broadening and halo structure. The intensity for this light source was adjusted to match observations (NR event topologies measured in the MIGDAL experiment) and represents 15$\%$ of the intensity observed from light that travels directly from GEM to lens, integrated over the entire simulated GEM surface. This represents less than $0.5\%$ of the photons generated in charge avalanche. Without accurate simulations of MIGDAL reflections, the results in Section \ref{sec:Discussion} provide both a minimum modeled increase that is based on Geant4 GEM structure optical redistribution and an example of the change from including reflective surfaces beyond the GEM. We acknowledge the reflections are not accurately modeled and can lead to large differences in the changes to particle tracks. The results should not be taken as the correct MH-PSF expected for MIGDAL but as a demonstration of the importance for accurately modeling reflections.

\section{\label{sec:Discussion}Applying MH-PSF}

In Section \ref{sec:Results}, we showed that light produced in a single GEM hole travels through the substrate and out neighboring holes, a result which is qualitatively consistent with the MIGDAL experiment's observed ``halos" in its G-GEM-based optical readout
\cite{Schueler2025}. In Section \ref{sec:Simulations} we
modeled a G-GEM MH-PSF using Geant4. The simulation qualitatively reproduces the observed glowing-paint redistribution pattern, with improved agreement after including a phenomenological reflected-light component.
In the following, we quantify the effect of the MH-PSF on
downstream observables recorded by an OTPC by applying Geant4 two stacked G-GEM MH-PSF
to simulated particle tracks. 
This MH-PSF is distinct from both the paint-measured MH-PSF and the phenomenologically tuned single-hole simulations. It is generated using the two G-GEM geometry, with photon starting distributions derived from Garfield++ avalanche simulations across the two GEMs. The minimum modeled track broadening and intensity changes are predicted from Geant4 photon transport through the G-GEM substrates for this MIGDAL-like optical configuration. Additional broadening from reflected light is more detector-dependent and remains sensitive to the assumed reflective surfaces and reflected-light contribution. The reflected-light case should therefore be interpreted as an illustrative estimate rather than a detector-independent prediction.
Although our results are 
relevant for any application that requires accurate energy and track reconstruction in a GEM-based
OTPC, we also discuss their impact on Migdal effect
searches.
The resulting halos from the MH-PSF could have a large impact in the latter where the goal is to 
reconstruct an electron and nuclear recoil track emerging from a common interaction vertex. The large dynamic range in their (d$E$/d$x$) and the low energies predicted for the electron add to the challenge of making a definitive identification of 
the Migdal effect topology.

\subsection{\label{subsec:ParticleTracks}Particle Tracks}

The simulation sample used for these studies consists of approximately 4,900 \SI{5.9}{keV} ERs, 4,000 fluorine NRs, and 6,500 carbon NRs. The simulated NR energy spectra are uniformly distributed between 50 and \SI{450}{keV_r} for fluorine and 50 and \SI{700}{keV_r} for carbon, with the upper limits chosen to approximately match the elastic scattering endpoint energies of \SI{2.5}{MeV} neutrons in CF$_4$. 
NR energies are shown with the electron equivalent energy \SI{}{keV_{ee}}, being the fraction of NR energy \SI{}{keV_{r}} after quenching.
Each primary ER and NR track is drifted and amplified following the 
readout simulation procedure described in Ref.$\,$\cite{MIGDAL2023} and then binned into a 2048$\times$1152 grid of \SI{40}{\micro\meter}-pitch pixels to match the field of view of the ORCA-Quest CMOS camera readout of the MIGDAL experiment. Intensities are scaled so that $^{55}$Fe energy spectra match real measurements. 
Our detector simulation procedure does not include any simulation of scintillation light, so our simulated track images are avalanche-multiplied \textit{charge} binned to the dimensions and field of view of MIGDAL's camera readout. Tracks are then randomly translated to appear in different locations across the image plane, and noise is added following the procedure described in Ref.$\,$\cite{Schueler2025}.

\subsection{\label{subsec:ApplyingConvolutionProcedure}Convolution Procedure}

\begin{figure}[pos=htbp]
    \centering
    \begin{subfigure}[t]{0.55\textwidth}
        \centering
        \includegraphics[width=1\textwidth]{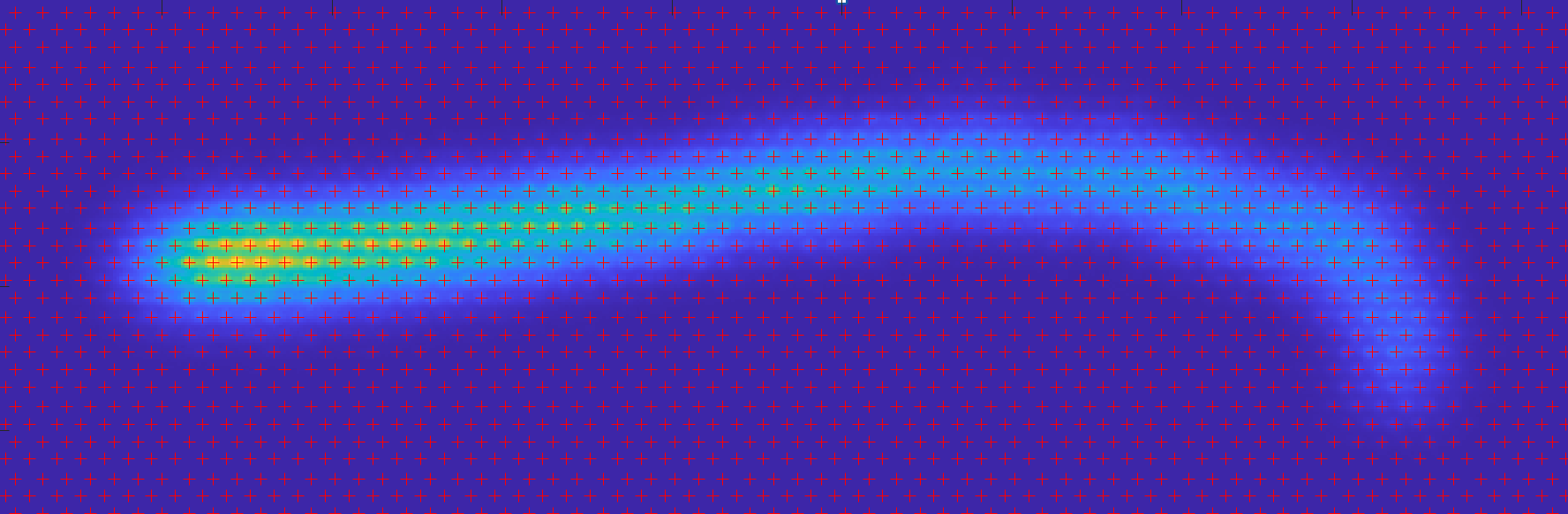}
        \caption{}
    \end{subfigure}
    \begin{subfigure}[t]{0.18\textwidth}
        \centering
        \includegraphics[width=1\textwidth]{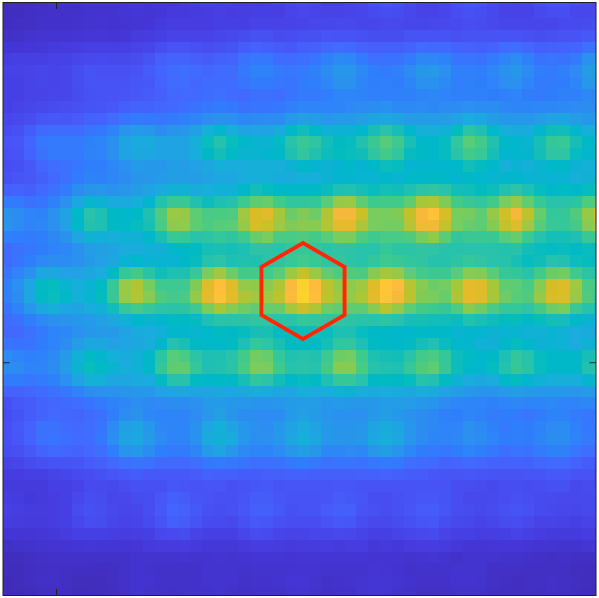}
        \caption{}
    \end{subfigure}
    \begin{subfigure}[t]{0.186\textwidth}
        \centering
        \includegraphics[width=1\textwidth]{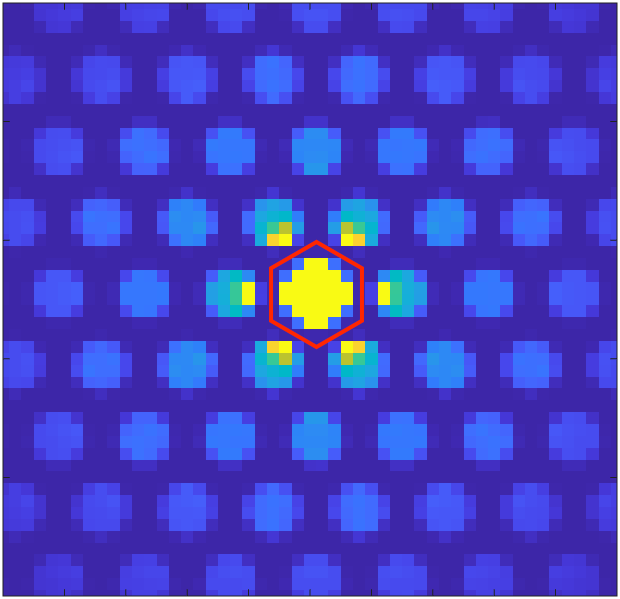}
        \caption{}
    \end{subfigure}
    \caption{Application of the simulated G-GEM MH-PSF to simulated particle tracks. (a) GEM-hole centers in the simulated particle-track image are identified using the known GEM pitch and hexagonal hole pattern. (b) A hexagonal region around each GEM-hole center is used to measure the particle-track intensity assigned to that hole. (c) The simulated G-GEM MH-PSF is normalized to the intensity within the corresponding hexagonal region, and the intensity outside the hexagon is added to model light redistributed into neighboring holes. 
    }
    \label{fig:DiscussionPhase1}
\end{figure}

To apply
our MH-PSF to track images, the first step is to locate all the GEM holes in the simulated images.\footnote{
\samepage When the simulated tracks were randomly translated, as described in the previous subsection, the new GEM hole locations were not saved.  
In principle, we could retrieve this information but we developed this technique to locate them as it would be required 
for tracks from real data.} 
We start with the brightest GEM hole, whose center is approximated using the location of the brightest pixel in the image. 
Next, we determine the GEM hole pitch in pixel units by counting the pixels between the visible holes in long particle tracks.
Using this and the hexagonal geometry of the GEM hole pattern, the remaining holes in the image can be found. The process is then iterative; overlaying the presumed GEM hole centers on tracks with visible GEM holes allows precise adjustments to the GEM hole pitch until the location of each hole is confidently identified. The results of this procedure are shown in Figure \ref{fig:DiscussionPhase1}a where the GEM hole centers are shown with red `+'s.

The image is then tiled into unit hexagonal cells, with each centered on a GEM hole (hexagon in Figure \ref{fig:DiscussionPhase1}b). 
With this, we define the intensity of a GEM hole {\it i}, in both the simulated track and MH-PSF\footnote{This requires the full 2D G-GEM MH-PSF, shown in Figure \ref{fig:DiscussionPhase1}c, with the same pixel-scale used for the simulated tracks.}, as the sum over the intensities of all pixels in their respective unit cells.
The MH-PSF is then scaled so that its central hole's intensity is equal to {\it I$_i$}, the intensity of GEM hole {\it i} in the simulated track. 

\begin{figure}[pos=htbp]
    \centering
    \begin{subfigure}[t]{0.32\textwidth}
        \centering
        \includegraphics[width=1\textwidth]{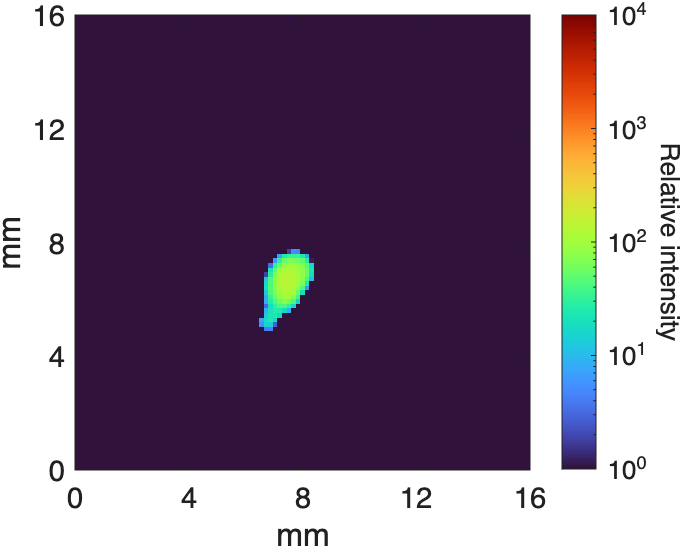}
    \end{subfigure}
    \begin{subfigure}[t]{0.32\textwidth}
        \centering
        \includegraphics[width=1\textwidth]{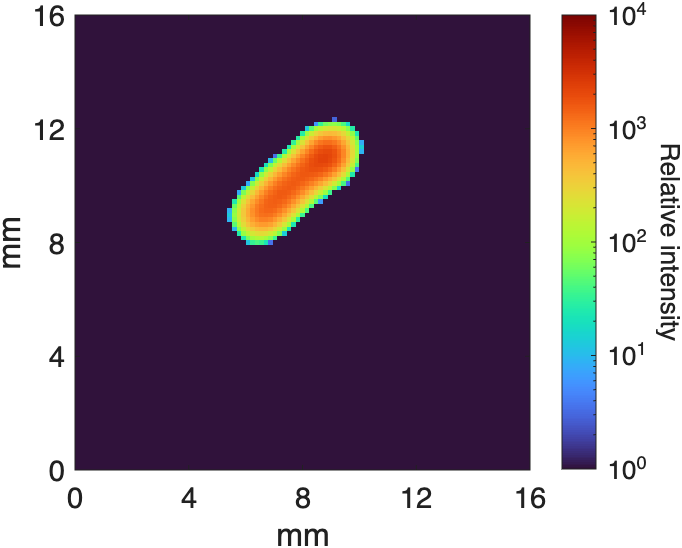}
        \captionsetup{labelformat=empty}
        \caption{Before simulated MH-PSF convolution}
    \end{subfigure}
    \begin{subfigure}[t]{0.32\textwidth}
        \centering
        \includegraphics[width=1\textwidth]{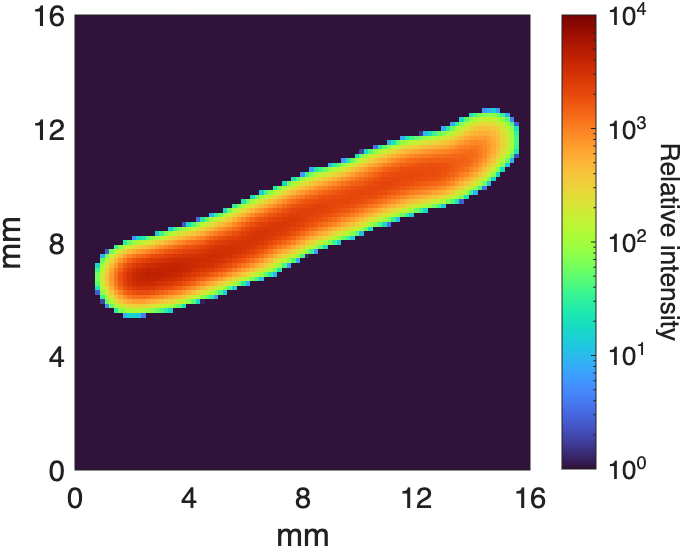}
    \end{subfigure}
    \begin{subfigure}[t]{0.32\textwidth}
        \centering
        \includegraphics[width=1\textwidth]{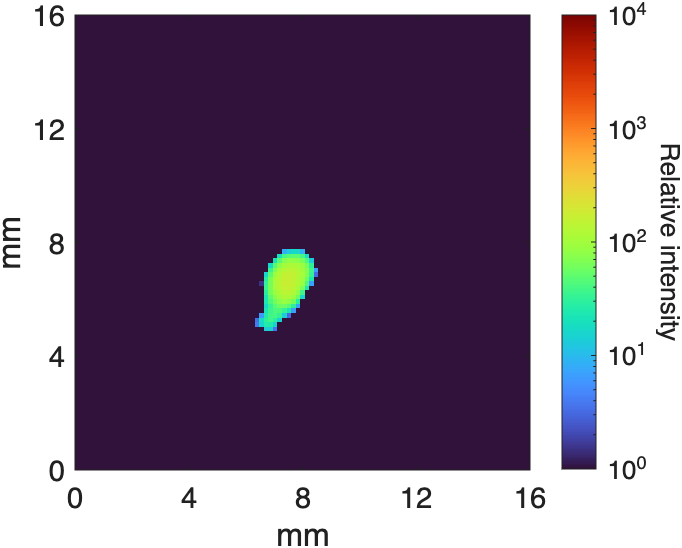}
    \end{subfigure}
    \begin{subfigure}[t]{0.32\textwidth}
        \centering
        \includegraphics[width=1\textwidth]{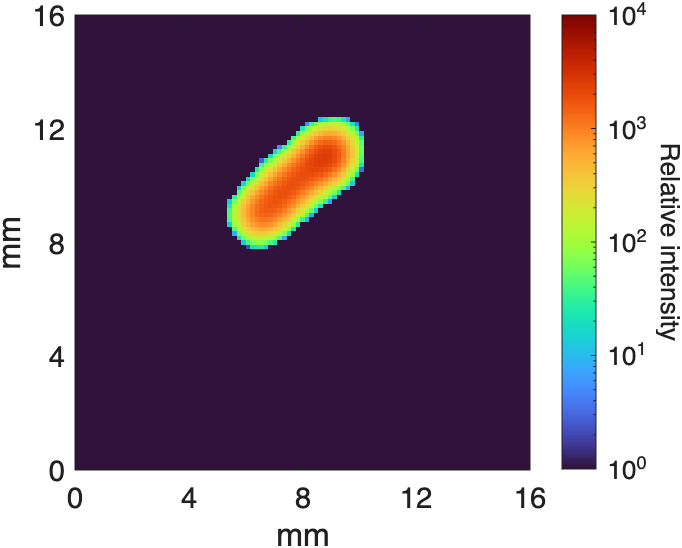}
        \captionsetup{labelformat=empty}
        \caption{After simulated MH-PSF convolution, minimum modeled with no reflections}
    \end{subfigure}
    \begin{subfigure}[t]{0.32\textwidth}
        \centering
        \includegraphics[width=1\textwidth]{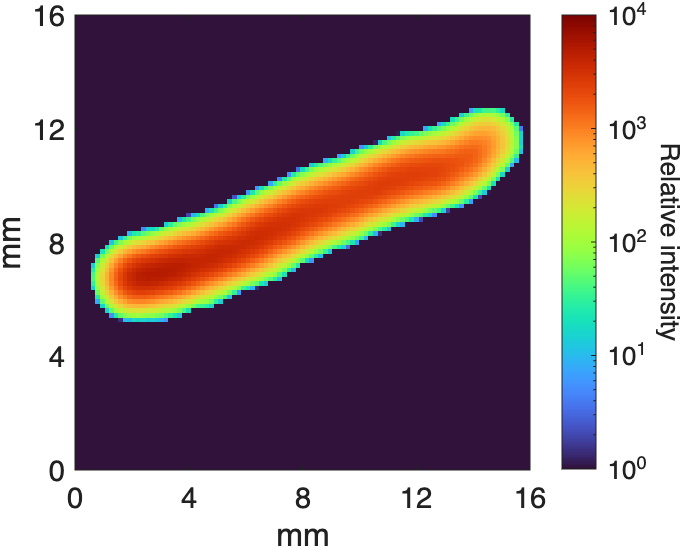}
    \end{subfigure}
        \begin{subfigure}[t]{0.32\textwidth}
        \centering
        \includegraphics[width=1\textwidth]{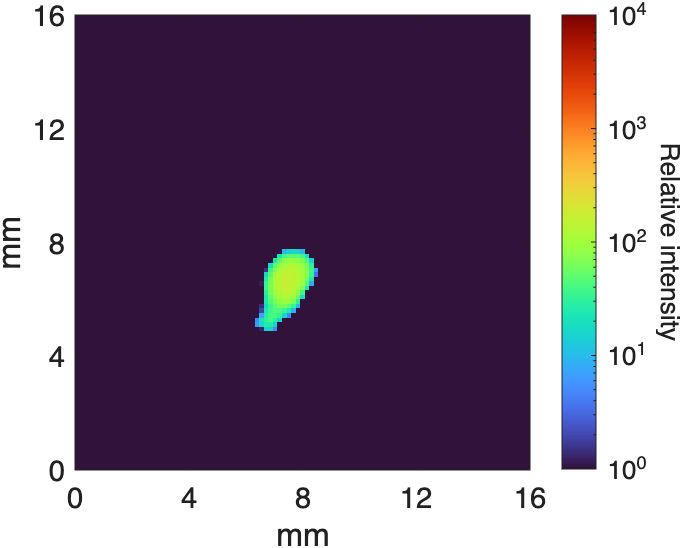}
    \end{subfigure}
    \begin{subfigure}[t]{0.32\textwidth}
        \centering
        \includegraphics[width=1\textwidth]{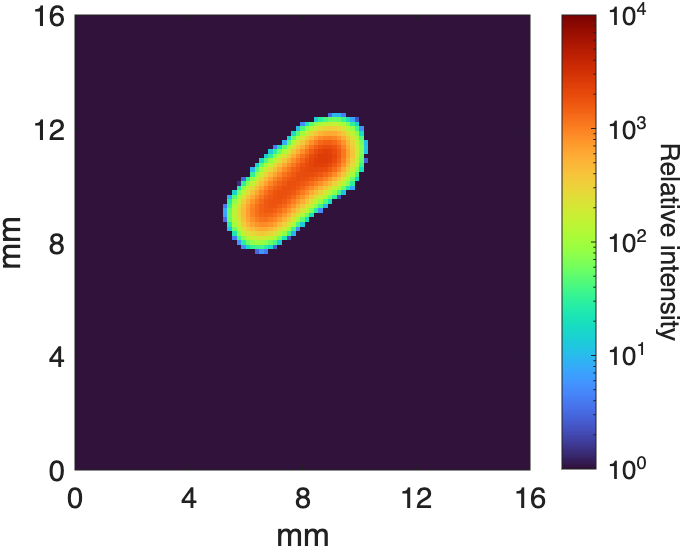}
        \captionsetup{labelformat=empty}
        \caption{After simulated MH-PSF convolution, with example reflections}
    \end{subfigure}
    \begin{subfigure}[t]{0.32\textwidth}
        \centering
        \includegraphics[width=1\textwidth]{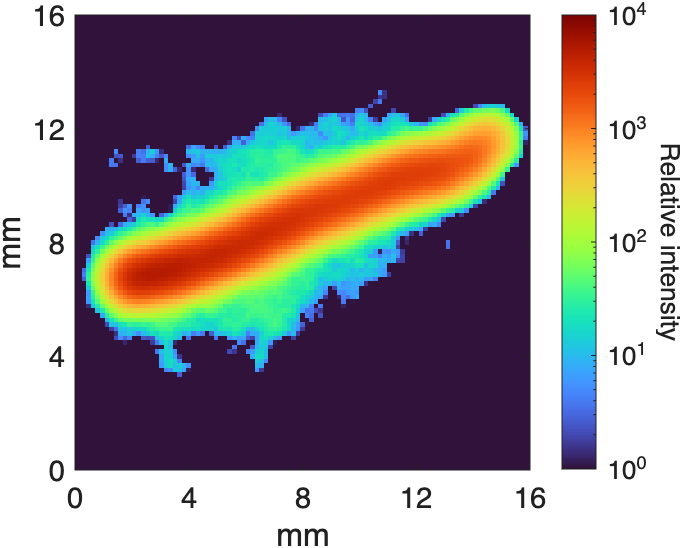}
    \end{subfigure}
    \caption{Simulated particle-track images before and after applying a simulated two-stacked G-GEM MH-PSF. The top row shows the original simulated tracks. The middle row shows the minimum modeled broadening case from Geant4 photon transport through the two G-GEM substrates, without additional reflective surfaces beyond the G-GEMs. The bottom row shows an illustrative case with an added reflected-light component to estimate possible additional broadening from surrounding reflective surfaces. Columns show a 5.9 keV electron recoil (left), a \(\SI{200}{keV_{ee}}\) fluorine nuclear recoil (center), and a \(\SI{600}{keV_{ee}}\) carbon nuclear recoil (right).}
    \label{fig:SimTracks}
\end{figure}

The intensity distribution in the region {\it outside} the central hole hexagon of the normalized MH-PSF now represents the halo contribution from {\it i}. 
The final step is to add this, pixel-by-pixel, to the simulated track by 

\begin{enumerate}
    \item Align the hexagons for GEM hole {\it i} and the MH-PSF.
    \item Per pixel {\it j} beyond the hexagons, take the normalized value from the MH-PSF to be the $\lambda$ parameter of a Poisson distribution.
   \item Generate a random integer from the Poisson distribution specified by the rate parameter $\lambda$. Poisson was chosen to convert fractional ADU values (after normalization) to integers and account for photon counting statistics.
   \item Add the integer to the intensity of pixel {\it j} in the simulated track. This process accounts for statistical fluctuations with the normalized intensity added per pixel {\it j} per GEM hole {\it i}.
   \item Repeat for each GEM hole {\it i}.
\end{enumerate}

Figure \ref{fig:SimTracks} provides examples of convolving realistic MH-PSFs, such as would be found in MIGDAL, with particle tracks. A minimum modeled increase that does not account for reflective surfaces beyond the G-GEM is shown, which is strictly derived from photons traveling through the GEM substrate directly towards the camera lens. Another increase is shown as an example of adding additional reflective surfaces as described in Section \ref{subsec:SimulationMIGDAL}. This example is not meant to be an accurate model for MIGDAL but to provide a demonstration of the importance for modeling reflections. The mean angle cut MH-PSF from each stacked GEM is used with the intensity contribution from each determined by Garfield++ simulations. The images had realistic noise added, were $4\times4$ binned, and Gaussian smoothed, and show the intensity above the noise threshold described in Section \ref{subsec:ApplyingParticleTracks}.

\subsection{\label{subsec:ApplyingParticleTracks}Applying MH-PSF Results}

\begin{figure}[pos=htbp]
    \centering
    \begin{subfigure}[t]{0.48\textwidth}
        \centering
        \includegraphics[width=1\textwidth]{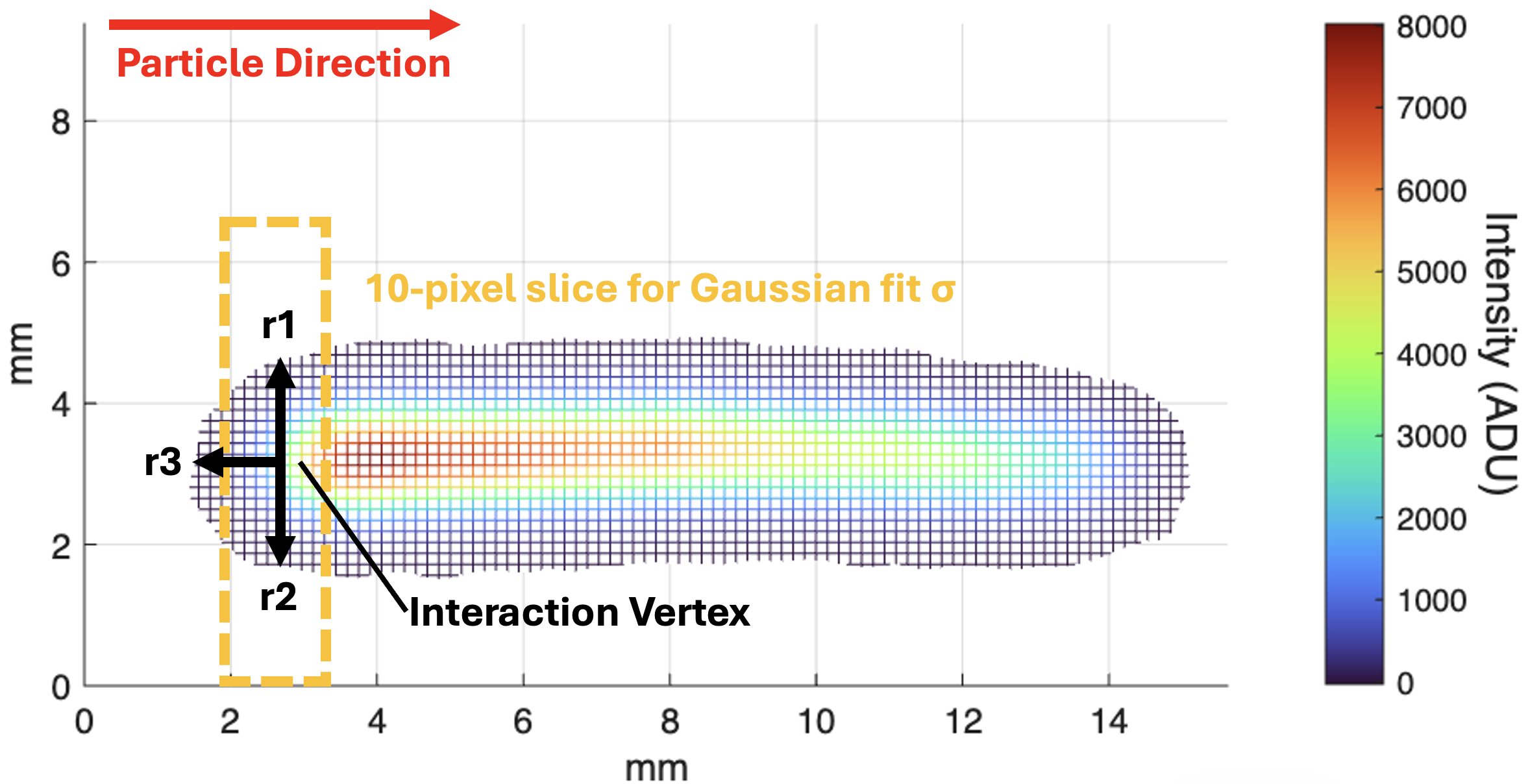}
    \end{subfigure}
    \caption{Measurements extracted from the simulated particle-track images. The measured quantities include total track intensity, number of pixels in the track, distances from the interaction vertex to the track edge (r1,r2,r3), and a slice through the interaction vertex used to fit the track width with a Gaussian and calculate the standard deviation \(\sigma\).}
    \label{fig:TrackMetrics}
\end{figure}

In the following, the term ``convolution" is used as a convenient shorthand for applying the MH-PSF to simulated images. The implemented procedure is not a strict mathematical convolution: instead of multiplying the image by a convolution kernel, it adds MH-PSF-weighted intensity contributions at the corresponding neighboring GEM hole locations. Thus, unconvolved and convolved, respectively, refer to images before and after the MH-PSF has been applied. 
Qualitatively, the differences in intensity and size of the track before and after applying the MH-PSF are apparent, as shown in Figure \ref{fig:SimTracks}.
To quantify the contributions of the optical halos on the track topology  and increased intensity of NRs and ERs, we define metrics that can be measured in simulated tracks both pre- and post-MH-PSF convolved and compare (see below and Figure \ref{fig:TrackMetrics}). 

\begin{figure}[pos=htbp]
    \centering
    \begin{subfigure}[t]{0.49\textwidth}
        \centering
        \includegraphics[width=1\textwidth]{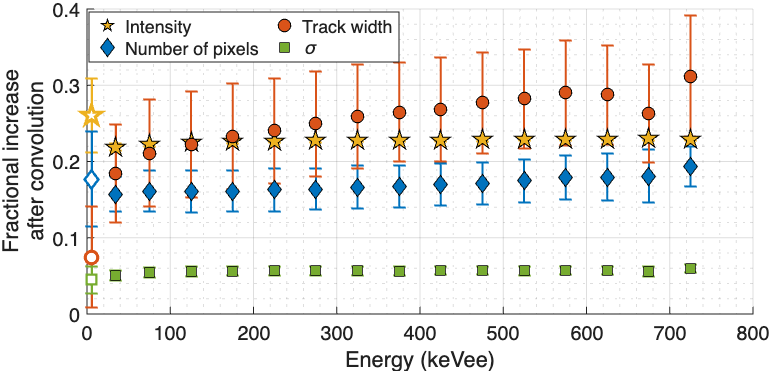}
        \caption{}
    \end{subfigure}
    \begin{subfigure}[t]{0.49\textwidth}
        \centering
        \includegraphics[width=1\textwidth]{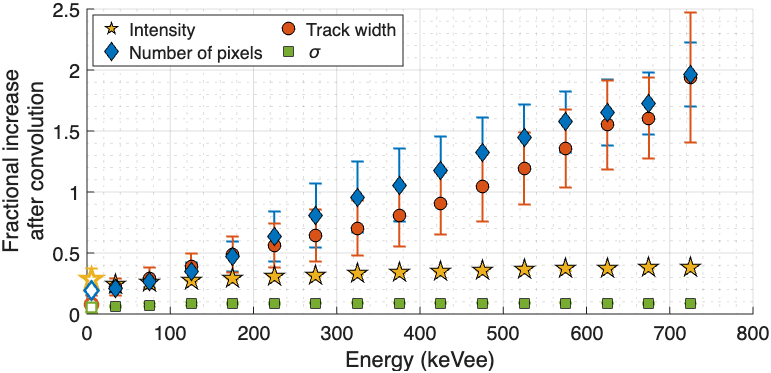}
        \caption{}
    \end{subfigure}
    \caption{Effects of the simulated two-stacked G-GEM MH-PSF on simulated particle-track observables. (a) Minimum modeled increase predicted from Geant4 photon transport through the G-GEM substrates, without additional reflective surfaces beyond the G-GEMs. (b) Illustrative case with an added reflected-light component to estimate possible additional changes from surrounding reflective surfaces. The 5.9 keV electron recoils are shown with hollow markers, and the nuclear recoils are shown with solid markers. Error bars indicate the standard deviation of the distribution at each point.}
    \label{fig:DiscussionPhase1-2}
\end{figure}

Comparisons of tracks without noise were performed to understand the true effect of the convolution process, which should result in a constant fractional increase in total intensity that is independent of energy or whether the particle is an ER or NR. That is what is observed where we see a constant $\sim23\%$ increase in total intensity (for the minimum modeled MH-PSF without reflections beyond the G-GEM). 
Figure \ref{fig:DiscussionPhase1-2}a shows the minimum modeled change in total intensity of particle tracks after applying the MH-PSF in the presence of realistic noise. Figure \ref{fig:DiscussionPhase1-2}b shows track changes with example reflections as described in Section \ref{subsec:SimulationMIGDAL}. The energy bin widths are from left: 0 for 5.9 keV ERs, \SI{35}{keV_{ee}} for the lowest energy NRs, \SI{50}{keV_{ee}} for all other NR bins. The fractional increase in intensity is calculated by the following equation:
\begin{equation} \label{eq:4x4Other}
    \text{Fractional X Increase} = \left\langle\frac{ X^C - X^U }{ X^U}\right\rangle.
\end{equation}
where ${X}$ can be the total intensity of a particle track, the number of pixels, r1, r2, r3, or $\sigma$ (see Figure \ref{fig:TrackMetrics}) from Convolved or Unconvolved images (superscript C and U).

Equation \ref{eq:4x4Other} is relevant for images taken in real experimental conditions where noise is included and the exact particle track boundaries are not known. Note that this equation essentially quantifies the difference in  what is observed in a GEM-based optically read out TPC (convolved term) versus one with a charge readout (unconvolved term). Both the convolved and unconvolved tracks include realistic noise. A $4\times4$ binning is applied to both images to improve the signal-to-noise ratio (SNR). For the purpose of calculating intensity, we sum inside a bounding box enclosing the track all pixels with a SNR $>$ 1.4 belonging to an identified track. Results using Equation \ref{eq:4x4Other} are shown in Figure \ref{fig:DiscussionPhase1-2}a (yellow pentagrams), where the total intensity of NRs increases from $\sim 22\%$ at low energies, and asymptotes to $\sim 23\%$ at high energies. The 5.9 keV ERs show an increase of $\sim26\%$.

Next we turn to the effects of the optical halos on geometrical track properties. For this we define metrics  that are related to track topology. The topological metrics used in this study measure the area and widths of NRs and ERs. The former is quantified using $N_{pix}$, the total number of pixels in the track. NR widths are measured at the interaction vertex (tail of NR track) and ER widths at the point of highest intensity (head of ER track). In order to characterize optical broadening, we are looking at points with high ionization density where the particle track can be observed. NR vertices and ER points of highest intensity both satisfy this criterion. While it would be ideal to consider ER interaction vertices, they are challenging to locate in a real experiment because of their low ionization density.
For NRs we define the width metrics as the distance from the vertex to the track boundary perpendicular to ($r1$ and $r2$), and anti-parallel to ($r3$) the particle track's direction (Figure \ref{fig:TrackMetrics}). For ERs the metrics are the same but relative to the point of highest intensity and assuming the particle direction based on drawing a straight line through the peak intensity and incrementally rotating until the maximum intensity along the line is found. 
The track boundary is where the SNR falls below 1.4. In addition, we also measure the effective diffusion ($\sigma$) of the particle tracks; since the addition of halos produces brighter and broader tracks, the net effect on $\sigma$ is unclear and needs to be quantified.
For this we take a 10 pixel wide strip centered on the NR vertex (point of highest intensity for ERs) shown in Figure \ref{fig:TrackMetrics}, project its intensity perpendicular to the track direction and fit a Gaussian to the resulting histogram. 

As expected, there is an increase in all topological quantities after adding the optical halos. The minimum modeled (Figure \ref{fig:DiscussionPhase1-2}a) fractional increase in the number of pixels, $N_{pix}$ (blue), for NRs starts at $\sim16\%$ and increases to $\sim19\%$ at high energies. 
5.9 keV ERs show a $\sim18\%$ increase.
The results for the fractional increase in $r1, r2$ and $r3$ were all similar, so only their mean is shown (red circles). For NRs it starts at $\sim18\%$ and rises to $\sim31\%$ at the highest energies, whereas for the 5.9 keV ERs the increase is $\sim7\%$. The lesser increase in track width for ERs than NRs, combined with the larger increase in intensity and $N_{pix}$, reveals optical halos do not broaden the ER peak intensity point similar to NR track width. This suggests the optical halos make more of the ER tail visible, bridging the interaction vertex to the peak intensity. The effective diffusion $\sigma$ of the tracks increases by $\sim4-6\%$ for all recoils. 

Figure \ref{fig:DiscussionPhase1-2}b shows an illustrative case in which reflective surfaces outside the G-GEM add significantly to the track width and $N_{pix}$, especially for higher-energy tracks. This occurs because reflected light can raise many pixels surrounding the particle tracks from just below the noise threshold to just above it. The quantitative broadening in this case depends strongly on the assumed reflectivity conditions, so these values should not be interpreted as detector-independent predictions. A detector-specific quantitative prediction would require measured or independently constrained reflectivity inputs for the MIGDAL optical surfaces. The reflective case is therefore included to demonstrate the potential importance of reflection modeling. 

Poisson fluctuations introduced by the MH-PSF application were evaluated by repeatedly applying the same MH-PSF procedure to a single simulated track with independent Poisson sampling of the redistributed light. The resulting spread in the fractional increases in total intensity and geometrical track properties was smaller than the track-to-track variation shown by the error bars in Figure \ref{fig:DiscussionPhase1-2}. Therefore, the plotted uncertainties are dominated by track-to-track variation, and the Poisson contribution from the MH-PSF application is not shown as a separate uncertainty.

The final study in this section considers the impact of the halos on 
the reconstruction of multi-particle track topologies. The search for a direct observation of the rare Migdal effect is an ideal example because it illustrates several challenges that are compounded by the halos. 
The identification of a Migdal event requires the unambiguous detection of a very faint, low-energy ER sharing a common interaction vertex with a $\sim 2$ orders of magnitude brighter NR. For further details and images of simulated Migdal events see MIGDAL \cite{MIGDAL2023,Schueler2025,Schueler-2025}.
Its detection efficiency will likely suffer due to broadening of the NR tracks by the optical halos, an effect  that is exacerbated by the vertex being near the brightest portion of the NR track.
To quantify the effects of the halo on the Migdal search, we consider the reduction in a portion of the ER track free from overlap with NR track edge. We define the non-overlapping portion of the ER track as the distance to the head of the ER track (the point of highest intensity and where most likely to be detected) minus that to the edge of the NR track (mean of $r1, r2$ and $r3$, shown in Figure \ref{fig:TrackMetrics}), both taken with respect to the vertex. Implicit in this definition is that we are mostly considering ER and NR tracks pointing in opposite hemispheres.

\begin{figure}[pos=htbp]
    \centering
    \begin{subfigure}[t]{0.48\textwidth}
        \centering
        \includegraphics[width=1\textwidth]{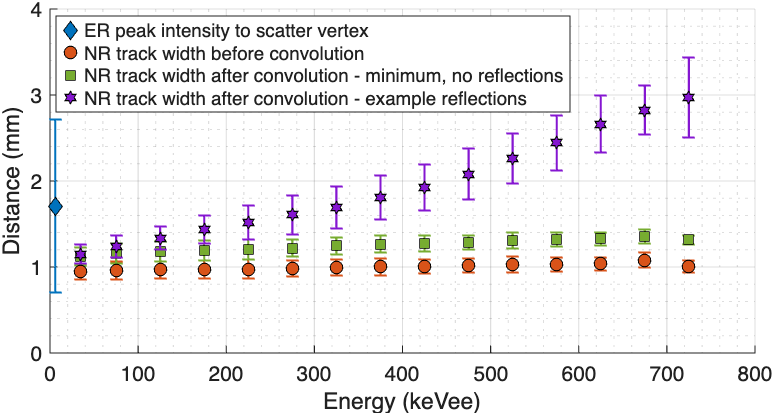}
    \end{subfigure}
    \caption{Implications of the simulated two-stacked G-GEM MH-PSF for Migdal-effect searches. The mean distance from the interaction vertex to the ER peak intensity is mostly unchanged after MH-PSF application, so only one blue diamond is shown. NR track widths change after MH-PSF application, with red circles showing the values before convolution, green squares showing the minimum modeled increase predicted from Geant4 photon transport through the G-GEM substrates without additional reflective surfaces, and purple hexagrams showing an illustrative case with an added reflected-light component to estimate possible additional changes from surrounding reflective surfaces. Increased NR track width can obscure more of the nearby ER signal, reducing the geometric separation between the ER and NR components. Error bars indicate the standard deviation of the distribution at each point.}
    \label{fig:DiscussionPhase1-3}
\end{figure}

The results are shown in Figure \ref{fig:DiscussionPhase1-3} using the same simulation sample and energy bin widths as above. The mean head-to-vertex distance for 5.9 keV ERs (blue diamond) is $\sim1.7$ mm with standard deviation of the distribution equal to $1.0$ mm, the large spread being due to ERs undergoing considerable straggling from multiple scattering. The peak intensity is mostly unchanged after convolution, just a slight broadening around the peak is observed. The NR track-edge-to-vertex distances range between $\sim0.95-1.1$ mm from low to high energies for the unconvolved tracks (red circles), growing to $\sim1.15-1.35$ mm after convolving the tracks with the MH-PSF (green squares). 
This represents a 27--42$\%$ reduction in the geometric separation between the NR boundary and the ER peak-intensity point, where the ER is most likely to be observed. The separation decreases from $0.75$ mm before convolution to $0.55$ mm for low-energy NRs and $0.35$ mm for high-energy NRs after convolution. Also shown are the NR track width changes with example reflections (purple hexagrams) as described in Section \ref{subsec:SimulationMIGDAL}.
This could present a significant loss in efficiency for a Migdal effect search, especially given that the predicted probabilities for the occurrence of the effect
fall exponentially with electron energy \cite{Ibe_2018}. A detailed estimate of this efficiency loss is beyond the scope of this work; nevertheless, the optical halos will impact MIGDAL and other applications where high resolution track reconstruction is required to identify complex, multi-particle track topologies.

\section{\label{sec:Summary}Discussion and Summary}

We developed a method for isolating a single GEM hole and filling it with a phosphorescent glow-in-the-dark paint to test the hypothesis that light produced within a GEM hole will propagate through the substrate and exit neighboring holes. Our method provides the requisite light source to measure a MH-PSF from light originating in a single GEM hole, enabling us to quantify the effects of light propagation through neighboring GEM holes on downstream observables in OTPCs. Our results confirm that, for the GEM materials and geometries tested here, light produced within a given hole can propagate through the GEM substrate and exit neighboring holes. The specific MH-PSF depends on the GEM geometry and materials used, with G-GEMs like those used in MIGDAL showing the greatest intensity from neighboring holes of all tested GEMs.

GEM optical effects can have a significant impact on OTPC measurements, and our work demonstrates that this effect is real and can be quantified. However, paint within a GEM hole will lead to a light source with a very different distribution than a charge avalanche induced scintillation. We developed more accurate MH-PSFs using Geant4 simulations of optical photons inside a GEM.
These simulations show that the MH-PSF observed in experiment will depend not only on the GEM substrate and geometry as shown in Section \ref{sec:Results}, but also on the distribution of photons emitted inside a GEM hole, the angles between the GEM hole and camera lens, and estimates of reflections in a given experiment as shown in Section \ref{sec:Simulations}. We found simulations are able to qualitatively reproduce the observed glowing-paint redistribution pattern using 2 parameters; the fraction of a GEM hole length filled with a cylinder of paint, and the fraction of uniform background light to represent reflections. We also developed more realistic models using two stacked G-GEMs and light distributions following Garfield++ simulations of charge avalanche. These detector-specific MH-PSF estimates can later be adapted to create deconvolution kernels, once the dominant optical redistribution mechanisms are sufficiently constrained for the detector configuration.

In Section \ref{sec:Discussion} we convolved our simulated G-GEM MH-PSF with simulated particle tracks representative of those observed in the MIGDAL detector. Doing so quantifies the effect of the G-GEM MH-PSF applied to tracks, providing necessary context to understand the optical broadening that will occur with G-GEMs and an optical readout. These studies show significant effects, with a minimum modeled increase in the intensity of tracks up to $26\%$, track widths increasing by up to $31\%$, and effective diffusion $\sigma$ increasing up to $\sim6\%$. We also show how these optical effects will impact experiments such as MIGDAL. When stitching the vertices of simulated ERs and NRs together, we show that these optical effects reduce the geometric separation between the NR boundary and the peak intensity of the ER track (where the ER is most likely to be observed) by 27-42$\%$. We therefore expect GEM optical effects to reduce the detection efficiency in rare-event searches, such as the Migdal effect, as the broadening of the track size will obscure more Migdal events. The magnitude of this reduction should not be interpreted as universal for all GEM-based OTPCs, since it depends on the detector-specific MH-PSF and optical configuration.

The present work uses a localized MH-PSF approach rather than a full detector-scale optical transport simulation. This choice allows the multi-hole redistribution effect to be isolated and applied to simulated images with manageable computational cost. A more complete detector-specific treatment would require full-detector optical simulations that track photon histories through the active GEM area and surrounding detector geometry. Such simulations can separate photons that travel directly from the scintillation region to the lens from photons that propagate through the GEM substrate and exit neighboring holes, or reflect from detector surfaces. This type of photon-history-resolved simulation is computationally expensive when performed track-by-track, but it provides a natural path toward constructing position-dependent MH-PSFs across the full imaging plane. Additional detector-specific effects, such as detector surface roughness, wavelength-dependent material properties, gas interactions, lens alignment, and camera response, could then be included to refine the MH-PSF. Developing a reliable deconvolution method therefore first requires a validated forward model capable of reproducing the relevant detector-specific optical effects. Only after such a full-detector optical model is established can spatially varying MH-PSFs be generated across the imaging plane and used as inputs for detector-specific deconvolution or image-correction methods. We therefore regard the localized MH-PSF treatment presented here as an experimentally constrained first step toward full detector-scale optical-response modeling.

Deconvolution methods have not yet been established, but they would involve solving the inverse problem associated with the MH-PSF application procedure described in Section \ref{subsec:ApplyingConvolutionProcedure}. In practice, the GEM hole positions could be determined by imaging the GEM plane before data collection, allowing the GEM pitch to be measured and a hexagonal GEM hole-to-pixel map to be constructed. A position-dependent library of MH-PSFs across the imaging plane could then provide the detector-specific response functions needed for image correction. However, deconvolution introduces additional challenges beyond forward application of the MH-PSF. In the forward application, the particle-track intensity assigned to each GEM hole is already known, so the MH-PSF can be normalized to that intensity before adding redistributed light to neighboring holes. In real data, the measured intensity in a given GEM hole can include both local scintillation produced in that hole and redistributed light originating from nearby holes. Therefore, a deconvolution procedure must determine which GEM holes contained local scintillation and how much of the measured intensity in those holes originates from local scintillation rather than from light redistributed from neighboring holes.

In the three classes of GEMs tested, our measurements of a MH-PSF confirm that tracks observed in an OTPC will be enhanced and broadened relative to those observed in a charge readout or predicted from simulations following electrons during charge avalanche. The magnitude of this effect is not universal and is expected to depend on the scintillation spatial distribution, GEM material, hole geometry, and the surrounding detector geometry that contribute to reflections and lens acceptance. An additional dependence of unknown magnitude is expected on the surface roughness, the gas mixture and pressure, the scintillation spectrum, the wavelength-dependent refractive indices, and the quantum efficiency of the camera. Experiments with the goal of creating the highest resolution particle tracks should consider these effects. Future GEM engineering and manufacturing may consider opacity and geometry changes to minimize the MH-PSF. Experiments should be aware of GEM optical effects, and for best results, model these effects for potential deconvolution methods.


\section*{Acknowledgements}
This work has been supported by the U.S. Department of Energy, Office of Science, Office of High Energy Physics, under Award Number DE-SC0022357; by the U.S. National Science Foundation under Award number 2209307, by the UKRI’s Science $\&$ Technology Facilities Council through the Grant Number ST/X508913/1; ET acknowledges the Graduate Instrumentation Research Award funded by the U.S. Department of Energy, Office of Science, Office of High Energy Physics. We are grateful to members of the MIGDAL Collaboration for discussions. For the purpose of open access, the authors have applied a Creative Commons Attribution (CC BY) license to any Author Accepted Manuscript arising from this submission.

\section*{Data Availability}

The data that support the findings of this study are available from the corresponding author upon reasonable request.

\section*{Declaration of Competing Interest}

The authors declare that they have no known competing financial interests or personal relationships that could have appeared to influence the work reported in this paper.





\printcredits

\bibliographystyle{elsarticle-num}

\bibliography{cas-refs}

@article{MIGDAL2023,
title = "{The MIGDAL experiment: Measuring a rare atomic process to aid the search for dark matter}",
journal = {Astroparticle Physics},
volume = {151},
pages = {102853},
year = {2023},
issn = {0927-6505},
doi = {10.1016/j.astropartphys.2023.102853},
author = {H.M. Araújo and S.N. Balashov and J.E. Borg  and others},
collaboration = {MIGDAL Collaboration},
old_author = {H.M. Araújo and S.N. Balashov and J.E. Borg and F.M. Brunbauer and C. Cazzaniga and C.D. Frost and F. Garcia and A.C. Kaboth and M. Kastriotou and I. Katsioulas and A. Khazov and H. Kraus and V.A. Kudryavtsev and S. Lilley and A. Lindote and D. Loomba and M.I. Lopes and E. Lopez Asamar and P. Luna Dapica and P.A. Majewski and T. Marley and C. McCabe and A.F. Mills and M. Nakhostin and T. Neep and F. Neves and K. Nikolopoulos and E. Oliveri and L. Ropelewski and E. Tilly and V.N. Solovov and T.J. Sumner and J. Tarrant and R. Turnley and M.G.D. {van der Grinten} and R. Veenhof}
}

@article{Brunbauer2018,
doi = {10.1088/1748-0221/13/02/T02006},
year = {2018},
month = {feb},
publisher = {},
volume = {13},
number = {02},
pages = {T02006},
author = {Brunbauer, Florian M. and Lupberger, M. and Oliveri, E. and others},
old_author = {Brunbauer, F.M. and Lupberger, M. and Oliveri, E. and Resnati, F. and Ropelewski, L. and Streli, C. and Thuiner, P. and Stenis, M. van},
title = "{Radiation imaging with optically read out GEM-based detectors}",
journal = {Journal of Instrumentation}
}

@ARTICLE{Brunbauer2018-2,
  author={Brunbauer, Florian M. and Garcia, Francisco and Korkalainen, Tero and others},
old_author={Brunbauer, Florian M. and Garcia, Francisco and Korkalainen, Tero and Lugstein, Alois and Lupberger, Michael and Oliveri, Eraldo and Pfeiffer, Dorothea and Ropelewski, Leszek and Thuiner, Patrik and Schinnerl, Markus},
  journal={IEEE Transactions on Nuclear Science}, 
  title="{Combined Optical and Electronic Readout for Event Reconstruction in a GEM-Based TPC}", 
  year={2018},
  volume={65},
  number={3},
  pages={913-918},
  doi={10.1109/TNS.2018.2800775}}

@article{Phan2020,
doi = {10.1088/1748-0221/15/05/P05012},
year = {2020},
month = {may},
publisher = {},
volume = {15},
number = {05},
pages = {P05012},
author = {Phan, N.S. and Lee, E.R. and Loomba, D.},
title = "{Imaging 55Fe electron tracks in a GEM-based TPC using a CCD readout}",
journal = {Journal of Instrumentation}
}

@article{TAKAHASHI2013,
title = "{Development of a glass GEM}",
journal = {Nuclear Instruments and Methods in Physics Research Section A: Accelerators, Spectrometers, Detectors and Associated Equipment},
volume = {724},
pages = {1-4},
year = {2013},
issn = {0168-9002},
doi = {10.1016/j.nima.2013.04.089},
author = {Hiroyuki Takahashi and Yuki Mitsuya and Takeshi Fujiwara and Takashi Fushie}
}

@article{FRAGA2001125,
title = "{Optical readout of GEMs}",
journal = {Nuclear Instruments and Methods in Physics Research Section A: Accelerators, Spectrometers, Detectors and Associated Equipment},
volume = {471},
number = {1},
pages = {125-130},
year = {2001},
note = {Imaging 2000},
issn = {0168-9002},
doi = {10.1016/S0168-9002(01)00972-X},
author = {F.A.F Fraga and L.M.S Margato and S.T.G Fetal and others},
old_author = {F.A.F Fraga and L.M.S Margato and S.T.G Fetal and M.M.F.R Fraga and R {Ferreira Marques} and A.J.P.L Policarpo}
}

@article{PINCI2019453,
title = "{High resolution TPC based on optically readout GEM}",
journal = {Nuclear Instruments and Methods in Physics Research Section A: Accelerators, Spectrometers, Detectors and Associated Equipment},
volume = {936},
pages = {453-455},
year = {2019},
note = {Frontier Detectors for Frontier Physics: 14th Pisa Meeting on Advanced Detectors},
issn = {0168-9002},
doi = {10.1016/j.nima.2018.11.085},
author = {D. Pinci and E. Baracchini and G. Cavoto and E. {Di Marco} and others},
old_author = {D. Pinci and E. Baracchini and G. Cavoto and E. {Di Marco} and M. Marafini and G. Mazzitelli and F. Renga and S. Tomassini and C. Voena}
}

@article{Schueler2025,
  title = "{Transforming a rare event search into a not-so-rare event search in real-time with deep learning-based object detection}",
  author = {Schueler, J. and Ara\'ujo, H. M. and Balashov, S. N. and others},
old_author = {Schueler, J. and Ara\'ujo, H. M. and Balashov, S. N. and Borg, J. E. and Brew, C. and Brunbauer, F. M. and Cazzaniga, C. and Cottle, A. and Frost, C. D. and Garcia, F. and Hunt, D. and Kaboth, A. C. and Kastriotou, M. and Katsioulas, I. and Khazov, A. and Knights, P. and Kraus, H. and Kudryavtsev, V. A. and Lilley, S. and Lindote, A. and Lisowska, M. and Loomba, D. and Lopes, M. I. and Lopez Asamar, E. and Luna Dapica, P. and Majewski, P. A. and Marley, T. and McCabe, C. and Millins, L. and Mills, A. F. and Nakhostin, M. and Nandakumar, R. and Neep, T. and Neves, F. and Nikolopoulos, K. and Oliveri, E. and Ropelewski, L. and Solovov, V. N. and Sumner, T. J. and Tarrant, J. and Tilly, E. and Turnley, R. and Veenhof, R.},
  collaboration = {MIGDAL Collaboration},
  journal = {Phys. Rev. D},
  volume = {111},
  issue = {7},
  pages = {072004},
  numpages = {26},
  year = {2025},
  month = {Apr},
  publisher = {American Physical Society},
  doi = {10.1103/PhysRevD.111.072004}
}

@article{Nygren1978,
    author = {Marx, Jay N. and Nygren, David R.},
    title = "{The Time Projection Chamber}",
    journal = {Physics Today},
    volume = {31},
    number = {10},
    pages = {46-53},
    year = {1978},
    month = {10},
    issn = {0031-9228},
    doi = {10.1063/1.2994775}
}

@article{Vahsen2021,
   author = "Vahsen, Sven E. and O'Hare, Ciaran A.J. and Loomba, Dinesh",
   title = "{Directional Recoil Detection}", 
   journal= "Annual Review of Nuclear and Particle Science",
   year = "2021",
   volume = "71",
   number = "Volume 71, 2021",
   pages = "189-224",
   doi = "10.1146/annurev-nucl-020821-035016",
   publisher = "Annual Reviews",
   issn = "1545-4134",
   type = "Journal Article",
  }

@article{Morgan2003,
title = "{DRIFT: a directionally sensitive dark matter detector}",
journal = {Nuclear Instruments and Methods in Physics Research Section A: Accelerators, Spectrometers, Detectors and Associated Equipment},
volume = {513},
number = {1},
pages = {226-230},
year = {2003},
note = {Proceedings of the 6th International Conference on Position-Sensitive Detectors},
issn = {0168-9002},
doi = {10.1016/j.nima.2003.08.037},
author = {Ben Morgan}
}

@INPROCEEDINGS{Battat2009,
       author = {{Battat}, J.~B.~R. and {Allien}, S. and {Caldwell}, T. and others},
       old_author = {{Battat}, J.~B.~R. and {Allien}, S. and {Caldwell}, T. and {Dujmic}, D. and {Dushkin}, A. and {Fisher}, P. and {Golub}, F. and {Goyal}, S. and {Henderson}, S. and {Inglis}, A. and {Lanza}, R. and {Lopez}, J. and {Kaboth}, A. and {Kohse}, G. and {Monroe}, J. and {Sciolla}, G. and {Skvorodnev}, B.~N. and {Tomita}, H. and {Vanderspek}, R. and {Wellenstein}, H. and {Yamamoto}, R.},
        title = "{DMTPC: A dark matter detector with directional sensitivity}",
    booktitle = {10th Conference on the Intersections of Particle and Nuclear Physics},
         year = 2009,
       editor = {{Marshak}, Marvin L.},
       series = {American Institute of Physics Conference Series},
       volume = {1182},
        month = dec,
    publisher = {AIP},
        pages = {276-279},
          doi = {10.1063/1.3293799}
}

@article{Shimada2023,
    author = {Shimada, Takuya and Higashino, Satoshi and Ikeda, Tomonori and others},
old_author = {Shimada, Takuya and Higashino, Satoshi and Ikeda, Tomonori and Nakamura, Kiseki and Yakabe, Ryota and Hashimoto, Takashi and Ishiura, Hirohisa and Nakamura, Takuma and Nakazawa, Miki and Kubota, Ryo and Nakayama, Ayaka and Ito, Hiroshi and Ichimura, Koichi and Abe, Ko and Kobayashi, Kazuyoshi and Tanimori, Toru and Kubo, Hidetoshi and Takada, Atsushi and Sekiya, Hiroyuki and Takeda, Atsushi and Miuchi, Kentaro},
    title = "{Direction-sensitive dark matter search with 3D-vector-type tracking in NEWAGE}",
    journal = {Progress of Theoretical and Experimental Physics},
    volume = {2023},
    number = {10},
    pages = {103F01},
    year = {2023},
    month = {10},
    issn = {2050-3911},
    doi = {10.1093/ptep/ptad120}
}

@article{Santos_2011,
   title="{MIMAC: A micro-tpc matrix for directional detection of dark matter}",
   volume={309},
   ISSN={1742-6596},
   DOI={10.1088/1742-6596/309/1/012014},
   journal={Journal of Physics: Conference Series},
   publisher={IOP Publishing},
   author={Santos, D and Billard, J and Bosson, G and others},
   old_author={Santos, D and Billard, J and Bosson, G and Bouly, J L and Bourrion, O and Fourel, Ch and Grignon, C and Guillaudin, O and Mayet, F and Richer, J P and Delbart, A and Ferrer, E and Giomataris, I and Iguaz, F J and Mols, J P and Golabek, C and Lebreton, L},
   year={2011},
   month=aug, pages={012014} }

@article{Amaro_2023,
   title="{The CYGNO experiment, a directional detector for direct Dark Matter searches}",
   volume={1054},
   ISSN={0168-9002},
   DOI={10.1016/j.nima.2023.168325},
   journal={Nuclear Instruments and Methods in Physics Research Section A: Accelerators, Spectrometers, Detectors and Associated Equipment},
   publisher={Elsevier BV},
   author={Amaro, Fernando Domingues and Baracchini, Elisabetta and Benussi, Luigi and others},
old_author={Amaro, Fernando Domingues and Baracchini, Elisabetta and Benussi, Luigi and Bianco, Stefano and Capoccia, Cesidio and Caponero, Michele and Cardoso, Danilo Santos and Cavoto, Gianluca and Cortez, André and Costa, Igor Abritta and Dané, Emiliano and Dho, Giorgio and Di Giambattista, Flaminia and Di Marco, Emanuele and D’Imperio, Giulia and Iacoangeli, Francesco and Lima, Herman Pessoa and Lopes, Guilherme Sebastiao Pinheiro and Maccarrone, Giovanni and Mano, Rui Daniel Passos and Gregorio, Robert Renz Marcelo and Marques, David José Gaspar and Mazzitelli, Giovanni and McLean, Alasdair Gregor and Messina, Andrea and Monteiro, Cristina Maria Bernardes and Nobrega, Rafael Antunes and Pains, Igor Fonseca and Paoletti, Emiliano and Passamonti, Luciano and Pelosi, Sandro and Petrucci, Fabrizio and Piacentini, Stefano and Piccolo, Davide and Pierluigi, Daniele and Pinci, Davide and Prajapati, Atul and Renga, Francesco and da Cruz Roque, Rita Joanna and Rosatelli, Filippo and Russo, Alessandro and dos Santos, Joaquim Marques Ferreira and Saviano, Giovanna and Spooner, Neil John Curwen and Tesauro, Roberto and Tommasini, Sandro and Torelli, Samuele},
   year={2023},
   month=sep, pages={168325} }

@article{Wang2020,
title = "{PandaX-III high pressure xenon TPC for Neutrinoless Double Beta Decay search}",
journal = {Nuclear Instruments and Methods in Physics Research Section A: Accelerators, Spectrometers, Detectors and Associated Equipment},
volume = {958},
pages = {162439},
year = {2020},
note = {Proceedings of the Vienna Conference on Instrumentation 2019},
issn = {0168-9002},
doi = {10.1016/j.nima.2019.162439},
author = {Shaobo Wang}
}

@article{Sokolowsak2024,
  title = "{Decay study of $^{11}\mathrm{Be}$ with an optical time-projection chamber}",
  author = {Soko\l{}owska, N. and Guadilla, V. and Mazzocchi, C. and others},
old_author = {Soko\l{}owska, N. and Guadilla, V. and Mazzocchi, C. and Ahmed, R. and Borge, M. J. G. and Cardella, G. and Ciemny, A. A. and Cosentino, L. G. and De Filippo, E. and Fedosseev, V. and Fija\l{}kowska, A. and Fraile, L. M. and Geraci, E. and Giska, A. and Gnoffo, B. and Granados, C. and Janas, Z. and Janiak, \L{}. and Johnston, K. and Kami\ifmmode \acute{n}\else \'{n}\fi{}ski, G. and Korgul, A. and Kubiela, A. and Maiolino, C. and Marsh, B. and Martorana, N. S. and Miernik, K. and Molkanov, P. and Ovejas, J. D. and Pagano, E. V. and Pirrone, S. and Pomorski, M. and Quynh, A. M. and Riisager, K. and Russo, A. and Russotto, P. and \ifmmode \acute{S}\else \'{S}\fi{}wiercz, A. and Vi\~nals, S. and Wilkins, S. and Pf\"utzner, M.},
  collaboration = {ISOLDE Collaboration},
  journal = {Phys. Rev. C},
  volume = {110},
  issue = {3},
  pages = {034328},
  numpages = {16},
  year = {2024},
  month = {Sep},
  publisher = {American Physical Society},
  doi = {10.1103/PhysRevC.110.034328}
}

@ARTICLE{Black2007,
       author = {{Black}, J.~K. and {Baker}, R.~G. and {Deines-Jones}, P. and others},
       old_author = {{Black}, J.~K. and {Baker}, R.~G. and {Deines-Jones}, P. and {Hill}, J.~E. and {Jahoda}, K.},
        title = "{X-ray polarimetry with a micropattern TPC}",
      journal = {Nuclear Instruments and Methods in Physics Research A},
         year = 2007,
        month = nov,
       volume = {581},
       number = {3},
        pages = {755-760},
          doi = {10.1016/j.nima.2007.08.144}
}

@INPROCEEDINGS{Fiorina2024,
       author = {{Fiorina}, Davide and {Baracchini}, Elisabetta and {Dho}, Giorgio and others},
       old_author = {{Fiorina}, Davide and {Baracchini}, Elisabetta and {Dho}, Giorgio and {Soffitta}, Paolo and {Torelli}, Samuele and {Marques}, David J.~G. and {Di Giambattista}, Flaminia and {Prajapati}, Atul and {Costa}, Enrico and {Fabiani}, Sergio and {Muleri}, Fabio and {Di Marco}, Alessandro and {Mazzitelli}, Giovanni},
        title = "{HypeX: high yield polarimetry experiment in x-rays}",
    booktitle = {X-Ray, Optical, and Infrared Detectors for Astronomy XI},
         year = 2024,
       editor = {{Holland}, Andrew D. and {Minoglou}, Kyriaki},
       series = {Society of Photo-Optical Instrumentation Engineers (SPIE) Conference Series},
       volume = {13103},
        month = aug,
          eid = {1310318},
        pages = {1310318},
          doi = {10.1117/12.3021559}
}

@article{Xu2024,
  title = "{Search for the Migdal effect in liquid xenon with keV-level nuclear recoils}",
  author = {Xu, J. and Adams, D. and Lenardo, B. G. and others},
  old_author = {Xu, J. and Adams, D. and Lenardo, B. G. and Pershing, T. and Mannino, R. L. and Bernard, E. and Kingston, J. and Mizrachi, E. and Lin, J. and Essig, R. and Mozin, V. and Kerr, P. and Bernstein, A. and Tripathi, M.},
  journal = {Phys. Rev. D},
  volume = {109},
  issue = {5},
  pages = {L051101},
  numpages = {7},
  year = {2024},
  month = {Mar},
  publisher = {American Physical Society},
  doi = {10.1103/PhysRevD.109.L051101}
}

@article{Alme2010,
title = "{The ALICE TPC, a large 3-dimensional tracking device with fast readout for ultra-high multiplicity events}",
journal = {Nuclear Instruments and Methods in Physics Research Section A: Accelerators, Spectrometers, Detectors and Associated Equipment},
volume = {622},
number = {1},
pages = {316-367},
year = {2010},
issn = {0168-9002},
doi = {10.1016/j.nima.2010.04.042},
author = {J. Alme and Y. Andres and H. Appelshäuser and others},
old_author = {J. Alme and Y. Andres and H. Appelshäuser and S. Bablok and N. Bialas and R. Bolgen and U. Bonnes and R. Bramm and P. Braun-Munzinger and R. Campagnolo and P. Christiansen and A. Dobrin and C. Engster and D. Fehlker and Y. Foka and U. Frankenfeld and J.J. Gaardhøje and C. Garabatos and P. Glässel and C. {Gonzalez Gutierrez} and P. Gros and H.-A. Gustafsson and H. Helstrup and M. Hoch and M. Ivanov and R. Janik and A. Junique and A. Kalweit and R. Keidel and S. Kniege and M. Kowalski and D.T. Larsen and Y. Lesenechal and P. Lenoir and N. Lindegaard and C. Lippmann and M. Mager and M. Mast and A. Matyja and M. Munkejord and L. Musa and B.S. Nielsen and V. Nikolic and H. Oeschler and E.K. Olsen and A. Oskarsson and L. Osterman and M. Pikna and A. Rehman and G. Renault and R. Renfordt and S. Rossegger and D. Röhrich and K. Røed and M. Richter and G. Rueshmann and A. Rybicki and H. Sann and H.-R. Schmidt and M. Siska and B. Sitár and C. Soegaard and H.-K. Soltveit and D. Soyk and J. Stachel and H. Stelzer and E. Stenlund and R. Stock and P. Strmeň and I. Szarka and K. Ullaland and D. Vranic and R. Veenhof and J. Westergaard and J. Wiechula and B. Windelband}
}

@article{Abe2011,
title = "{The T2K experiment}",
journal = {Nuclear Instruments and Methods in Physics Research Section A: Accelerators, Spectrometers, Detectors and Associated Equipment},
volume = {659},
number = {1},
pages = {106-135},
year = {2011},
issn = {0168-9002},
doi = {10.1016/j.nima.2011.06.067},
author = {K. Abe and N. Abgrall and H. Aihara and others}
}

@manual{ehdlens,
  author       = {{EHD Imaging GmbH}},
  title        = "{EHD F0.85 Lenses}",
  year         = 2020,
  month        = {March},
  address      = {Zum Rennplatz 15, D‐49401 Damme, Germany},
  note         = {\url{https://www.ehd.de/wp-content/uploads/2024/05/EHD-F085-Lenses.pdf} (accessed April 2026)},
  organization = {EHD Imaging GmbH}
}

@article{GEANT4:2002zbu,
    author = "Agostinelli, S. and others",
    collaboration = "GEANT4",
    title = "{GEANT4 - A Simulation Toolkit}",
    reportNumber = "SLAC-PUB-9350, FERMILAB-PUB-03-339, CERN-IT-2002-003",
    doi = "10.1016/S0168-9002(03)01368-8",
    journal = "Nucl. Instrum. Meth. A",
    volume = "506",
    pages = "250--303",
    year = "2003"
}

@article{Ibe_2018,
   title="{Migdal effect in dark matter direct detection experiments}",
   volume={2018},
   ISSN={1029-8479},
   DOI={10.1007/jhep03(2018)194},
   number={3},
   pages={194},
   journal={Journal of High Energy Physics},
   publisher={Springer Science and Business Media LLC},
   author={Ibe, Masahiro and Nakano, Wakutaka and Shoji, Yutaro and Suzuki, Kazumine},
   year={2018},
   month=mar }

@article{PEG3refractive,
SerialNo={34},
author={Hatsuda, M. and Motomura, Y. and Hashimoto, K.},
year={2007},
title={{Micro-Fabrication Process and Products of Photosensitive Etching Glass "PEG 3"}},
journal={New Glass},
volume={22},
number={1},
pages={75--80},
note={{https://www.newglass.jp/mag/TITL/maghtml/84-pdf/+84-p075.pdf}. (Accessed April 2026)},
}

@article{CopperRefractive,
SerialNo={35},
author={Polyanskiy, M. N.},
year={2024},
title={{Refractiveindex.info database of optical constants}},
journal={Scientific Data},
volume={11},
pages={94},
doi={10.1038/s41597-023-02898-2},
}

@inproceedings{BrunbauerRD51_2018,
author = {Brunbauer,  Florian M. and others},
  title = "{High-speed optical GEM readout}",
  series = {RD51 Collaboration Meeting and the "MPGD Stability" workshop},
  address = {Munich, Germany},
  year = {2018},
  note = {\url{https://indico.cern.ch/event/709670/contributions/3036126/attachments/1671185/2680963/FastTiming_3DPrinting_Brunbauer.pdf} (accessed April 2026)}
}

@article{Allison_2016,
title = "{Recent developments in Geant4}",
journal = {Nuclear Instruments and Methods in Physics Research Section A: Accelerators, Spectrometers, Detectors and Associated Equipment},
volume = {835},
pages = {186-225},
year = {2016},
issn = {0168-9002},
doi = {10.1016/j.nima.2016.06.125},
author = {J. Allison and K. Amako and J. Apostolakis and others},
old_author = {J. Allison and K. Amako and J. Apostolakis and P. Arce and M. Asai and T. Aso and E. Bagli and A. Bagulya and S. Banerjee and G. Barrand and B.R. Beck and A.G. Bogdanov and D. Brandt and J.M.C. Brown and H. Burkhardt and Ph. Canal and D. Cano-Ott and S. Chauvie and K. Cho and G.A.P. Cirrone and G. Cooperman and M.A. Cortés-Giraldo and G. Cosmo and G. Cuttone and G. Depaola and L. Desorgher and X. Dong and A. Dotti and V.D. Elvira and G. Folger and Z. Francis and A. Galoyan and L. Garnier and M. Gayer and K.L. Genser and V.M. Grichine and S. Guatelli and P. Guèye and P. Gumplinger and A.S. Howard and I. Hřivnáčová and S. Hwang and S. Incerti and A. Ivanchenko and V.N. Ivanchenko and F.W. Jones and S.Y. Jun and P. Kaitaniemi and N. Karakatsanis and M. Karamitros and M. Kelsey and A. Kimura and T. Koi and H. Kurashige and A. Lechner and S.B. Lee and F. Longo and M. Maire and D. Mancusi and A. Mantero and E. Mendoza and B. Morgan and K. Murakami and T. Nikitina and L. Pandola and P. Paprocki and J. Perl and I. Petrović and M.G. Pia and W. Pokorski and J.M. Quesada and M. Raine and M.A. Reis and A. Ribon and A. {Ristić Fira} and F. Romano and G. Russo and G. Santin and T. Sasaki and D. Sawkey and J.I. Shin and I.I. Strakovsky and A. Taborda and S. Tanaka and B. Tomé and T. Toshito and H.N. Tran and P.R. Truscott and L. Urban and V. Uzhinsky and J.M. Verbeke and M. Verderi and B.L. Wendt and H. Wenzel and D.H. Wright and D.M. Wright and T. Yamashita and J. Yarba and H. Yoshida}
}

@ARTICLE{Brunbauer-2025, 
AUTHOR={Brunbauer, F. M.  and Amedo, P.  and Flöthner, K. J.  and others},
old_AUTHOR={Brunbauer, F. M.  and Amedo, P.  and Flöthner, K. J.  and Gonzalez Diaz, D.  and Janssens, D.  and Leardini, S.  and Lisowska, M.  and Müller, H.  and Oliveri, E.  and Orlandini, G.  and Pfeiffer, D.  and Ropelewski, L.  and Sauli, F.  and Samarati, J.  and Scharenberg, L.  and van Stenis, M.  and Veenhof, R. },
         
TITLE="{Primary and secondary scintillation of CF4-based mixtures in low-pressure gaseous detectors}",
        
JOURNAL={Frontiers in Detector Science and Technology},
        
VOLUME={Volume 3 - 2025},

YEAR={2025},

DOI={10.3389/fdest.2025.1561739},

ISSN={2813-8031}}

@misc{garfieldpp,
  author       = {H. Schindler},
  title        = "{Garfield++ User Guide}",
  year         = {2022},
  note = {\url{https://garfieldpp.web.cern.ch/} (accessed April 2026)},
}

@article{Schueler-2025,
SerialNo={36},
author={Schueler, J. and Ara{\'u}jo, H. M. and Balashov, S. N. and others},
year={2026},
title={{Overlap-aware segmentation for topological reconstruction of obscured objects}},
journal={Machine Learning: Science and Technology},
volume={7},
number={3},
pages={035046},
doi={10.1088/2632-2153/ae6e18},
}

@ARTICLE{Antonietti-2025,

AUTHOR={Antonietti, R. and Baracchini, E. and Benussi, L. and others},

TITLE="{The CYGNO experiment}",

JOURNAL={Il Nuovo Cimento C},

VOLUME={48},

NUMBER={3},

PAGES={95},

YEAR={2025},

DOI={10.1393/ncc/i2025-25095-0}

}

@ARTICLE{Roberts-2019,

AUTHOR={Roberts, A. and Svihra, P. and Al-Refaie, A. and others},

TITLE="{First demonstration of 3D optical readout of a TPC using a single photon sensitive Timepix3 based camera}",

JOURNAL={Journal of Instrumentation},

VOLUME={14},

NUMBER={06},

PAGES={P06001},

YEAR={2019},

DOI={10.1088/1748-0221/14/06/P06001}

}

@ARTICLE{Lowe-2020,

AUTHOR={Lowe, A. and Majumdar, K. and Mavrokoridis, K. and others},

TITLE="{Optical Readout of the ARIADNE LArTPC Using a Timepix3-Based Camera}",

JOURNAL={Instruments},

VOLUME={4},

NUMBER={4},

PAGES={35},

YEAR={2020},

DOI={10.3390/instruments4040035}

}

@article{Amaro_2026,
   title="{Modeling the light response of an optically readout GEM based TPC for the CYGNO experiment}",
   volume={86},
   DOI={10.1140/epjc/s10052-026-15318-7},
   journal={Eur. Phys. J. C},
   author={Amaro, Fernando Domingues and Antonietti, Rita and Baracchini, Elisabetta and others},
   year={2026},
   month=feb, pages={123} }

@article{Cools_2024,
   title="{Spatial resolution studies using point spread function extraction in optically read out Micromegas and GEM detectors}",
   volume={1069},
   DOI={10.1016/j.nima.2024.169933},
   journal={Nuclear Instruments and Methods in Physics Research Section A: Accelerators, Spectrometers, Detectors and Associated Equipment},
   author={Cools, A. and Ferrer-Ribas, E. and Papaevangelou, T. and others},
   year={2024},
   pages={169933} }



\end{document}